\documentclass[12pt]{article}
\begin{document}

\renewcommand{\thefootnote}{\fnsymbol{footnote}}
\begin{center}
{\LARGE\baselineskip0.9cm 
A Grassmann integral equation\\[1.5cm]}

{\large K. Scharnhorst\footnote[1]{E-mail:
{\tt scharnh@physik.hu-berlin.de}}\footnote[2]{
Present address: Vrije Universiteit Amsterdam,
Faculty of Exact Sciences, Divison of Physics and Astronomy, 
Department of Theoretical Physics,
De Boelelaan 1081, 1081 HV Amsterdam, The Netherlands,
e-mail: {\tt scharnh@nat.vu.nl}}
}\\[0.cm]

{\small Humboldt-Universit\"at zu Berlin

Institut f\"ur Physik

Invalidenstr.\ 110

10115 Berlin

Federal Republic of Germany}\\[1.cm]

\begin {abstract}
The present study introduces and investigates
a new type of equation which is called
{\it Grassmann integral equation} in analogy to integral equations
studied in real analysis. A Grassmann integral equation is an 
equation which involves Grassmann (Berezin) integrations and which is
to be obeyed by an unknown function over
a (finite-dimensional) Grassmann algebra ${\cal G}_m$ 
(i.e., a sought after element of the Grassmann algebra ${\cal G}_m$).
A particular type of Grassmann integral equations is explicitly
studied for certain low-dimensional Grassmann algebras. The 
choice of the equation under investigation is motivated by 
the effective action formalism of (lattice) quantum field theory.
In a very general setting, for the Grassmann algebras 
${\cal G}_{2n}$, $n = 2, 3, 4$, the 
finite-dimensional analogues of the generating functionals of the 
Green functions are worked out explicitly by solving a coupled
system of nonlinear matrix equations. 
Finally, by imposing the condition 
$G[\{\bar\Psi\},\{\Psi\} ]= G_0[\{\lambda\bar\Psi\},\{\lambda\Psi\} ] 
+ const.$,
$0<\lambda\in{\bf R}$ ($\bar\Psi_k$, $\Psi_k$, $k=1,\ldots , n$,
are the generators of the Grassmann algebra ${\cal G}_{2n}$),
between the finite-dimensional analogues
$G_0$ and $G$ of the (``classical'') action and effective action
functionals, respectively, 
a special Grassmann integral equation is being established
and solved which also is equivalent to a coupled system of nonlinear
matrix equations. If $\lambda\not=1$, solutions to this Grassmann
integral equation exist for $n=2$ (and consequently, also for 
any even value of $n$, specifically, for $n=4$)
but not for $n=3$. If $\lambda=1$, the considered 
Grassmann integral equation (of course)
has always a solution which corresponds to a Gaussian integral,
but remarkably in the case $n=4$ a further solution is found which
corresponds to a non-Gaussian integral.
The investigation sheds light on the structures to be met
for Grassmann algebras ${\cal G}_{2n}$ with arbitrarily chosen $n$.
\end{abstract}

\end{center}

\renewcommand{\thefootnote}{\arabic{footnote}}
\thispagestyle{empty}

\newpage

\section{Introduction}

The problem to be studied in the present paper is a purely 
mathematical one and one might arrive at it
within various research programmes in mathematics and its applications.
Our starting point will be (lattice) quantum field theory 
\cite{creu}-\cite{smit} 
and for convenience 
we will mainly use its terminology throughout the study
(incidentally, for a finite-dimensional problem!). 
However, one could
equally well rely on the terminology of 
statistical mechanics or probability theory throughout.
We will be interested in certain aspects of
differential calculus in Grassmann (Gra\ss mann) algebras
\cite{bere1} and in particular in Grassmann analogues 
to integral equations studied in real analysis which we will
call {\it Grassmann integral equations}. 
A Grassmann integral equation is an 
equation which involves Grassmann (Berezin) integrations and which is
to be obeyed by an unknown function over
a (finite-dimensional) Grassmann algebra ${\cal G}_m$ 
(i.e., a sought after element of the Grassmann algebra ${\cal G}_m$).
To the best of our
knowledge this problem is considered for the first time in this
paper. Of course, the following comment is due.
Bearing in mind that in a Grassmann algebra taking a
(Grassmann) derivative and an integral are equivalent operations
we could equally well denote any Grassmann integral equation
as a Grassmann differential equation. There is an 
extensive literature on supersymmetric extensions of differential 
equations. Corresponding research has been performed in areas such as
supersymmetric field theory (see, e.g., \cite{wein}, Vol.\ 3), 
superconformal field theory, the study of supersymmetric
integrable models (see, e.g., \cite{keto,kras}), and superanalysis
(for a review of the latter see the recent book by Khrennikov \cite{khre1}, 
in particular Chap.\ 2, and references therein). Only few mathematical 
references exist which treat pure Grassmann differential equations (understood
in the narrow sense, i.e., in a non-supersymmetric setting)
\cite{fryd}-\cite{knya}. 
In the physics literature, specifically in the quantum field theoretic
literature, such equations (in general, for infinite-dimensional 
Grassmann algebras) can be found in studies of purely fermionic
models by means of the Schwinger-Dyson equations\footnote{
For a general discussion of these equations see, e.g., 
\cite{vasi}, Chap.\ 1, \S 7. p.\ 72
(Sect.\ 1.7, p.\ 75 of the English translation), 
\cite{itzy}, Sect.\ 10.1, p.\ 475, \cite{rive,zinn2}; 
in the context of 
purely fermionic theories these equations can be found displayed,
e.g., in \cite{cvit2}, Sect.\ 4.C, Exercise 4.C.2, p.\ 54, \cite{laws,gura1}.}
or the Schr\"odinger representation 
(\cite{barn3,flor} and follow-up references citing these).
Within the framework of supersymmetric generalizations 
of conventional analysis, it is customary 
to consider all structures in strict analogy to real (complex) analysis.
Consequently, as we will be lead to the problem of Grassmann
integral equations from the corresponding problem in real
analysis the choice of this term should not lead to any 
objection. Incidentally, it might be interesting to note
that Khrennikov \cite{khre1} mentions (at the end of Chap.\ 2,
p.\ 102 [p.\ 106 of the English translation]) integral equations
(item 9.) among the subjects which have not yet been studied
in superanalysis.\\

Having characterized in general the subject of the present study 
we will now explain in somewhat greater detail the problem we 
are interested in and where it arises from. Our motivation 
for the present investigation derives from quantum field theory.
Quantum field theory is a rich subject with many facets and
is being studied on the basis of a number of approaches and
methods. For the present purpose, we 
rely on the functional integral approach to Lagrangian quantum 
field theory (see, e.g., \cite{vasi}, \cite{itzy}, Chap.\ 9, p.\ 425, \cite{rive}, \cite{wein}, Vol.\ I, Chap.\ 9, p.\ 376). To begin with, consider 
the theory of a scalar field $\phi$ in $k$-dimensional
Minkowski space-time. By the following equations one defines
generating functionals for various types of Green functions of this 
field (see, e.g., \cite{vasi}, \cite{itzy}, {\it loc.\ cit.}, 
\cite{rive}, \cite{zinn2}, Chap.\ 6, \cite{wein}, Vol.\ II,
Chap.\ 16, p.\ 63, ).
\begin{eqnarray}
\label{M1a}
Z[J] &=& C\ \int D\phi\ \ {\rm e}^{\displaystyle\ 
i\Gamma_0 [\phi]\ +\ i \int d^kx\ J(x) \phi (x)}\\[0.3cm]
\label{M1b}
W[J] &=& -i \ln Z[J]\\[0.3cm]
\label{M1c}
\Gamma [\bar\phi] &=& W[J] - \int d^kx\ J(x) \bar\phi (x)\\[0.3cm]
\label{M1d}
\bar\phi (x) &=& {\delta W[J]\over \delta J(x)}
\end{eqnarray}
From eq.\ (\ref{M1c}) one finds the relation
\begin{eqnarray}
\label{M2}
{\delta \Gamma [\bar\phi]\over \delta \bar\phi (x)} &=& -\ J(x)\ .
\end{eqnarray}
In eq.\ (\ref{M1a}), $\int D\phi$ denotes the 
(infinite-dimensional) functional integration over the scalar field $\phi$.
$Z[J]$ is the generating functional of the Green 
functions\footnote{
$Z[J]$ can be understood as the (infinite-dimensional) Fourier transform of 
$\exp i\Gamma_0 [\phi]$, \cite{domi}, \cite{vasi}, Chap.\ 1, \S 6,
Subsect.\ 7, p.\ 71 (Subsect.\ 1.6.7, p.\ 73 of the English translation).}, 
$W[J]$ is the generating functional of the connected Green functions
while the (first) Legendre transform $\Gamma [\bar\phi]$ of $W[J]$ 
is the generating functional of the one-particle-irreducible
(1PI) Green functions. 
$\Gamma_0 [\phi]$ is the so-called classical action of
the theory and $C$ some fixed normalization constant. 
$\Gamma [\bar\phi]$ is also called
the effective action of the theory and, in principle, any 
information one might ever be interested in can be derived from it.\\

Eq.\ (\ref{M1a}) defines a map,
$g_1: \Gamma_0 [\phi ]\longrightarrow Z[J]$, from the 
class of functionals called classical actions to the 
class of functionals $Z$. Furthermore, we have 
mappings, $g_2: Z[J]\longrightarrow W[J]$, (eq.\ (\ref{M1b})), and 
$g_3: W[J]\longrightarrow \Gamma [\bar\phi]$
(eq.\ (\ref{M1c})). These three maps together define a
map $g_3\circ g_2\circ g_1 = f: \Gamma_0 [\phi ]
\longrightarrow \Gamma [\bar\phi]$
from the set of so-called classical actions to the set of 
effective actions (We will call $f$ the `action map'.).
In general, the action map is mathematically not well-defined in quantum 
field theory due to the occurrence of ultraviolet divergencies
and one has to apply a regularization procedure for making
proper mathematical sense of the above equations. A widely applied
approach which is very natural from a mathematical point of
view consists in studying quantum field theory not on a space-time 
continuum but on a space-time lattice (see, e.g., 
\cite{creu}-\cite{smit}). 
The map $f$ can be represented by the following single equation
which can be derived from the eqs.\ (\ref{M1a})-(\ref{M1c}).
\begin{eqnarray}
\label{M3}
{\rm e}^{\displaystyle\ i \Gamma [\bar\phi]}\ &=& C\ 
\int D\phi\ \ {\rm e}^{\displaystyle\ i\Gamma_0 
[\phi + \bar\phi]\ +\ i \int d^kx\ J(x) \phi (x)}
\end{eqnarray}
$J(x)$ is given here by eq.\ (\ref{M2}), consequently, eq.\ (\ref{M3})
is only an implicit representation of the map $f$. For any quantum
field theory, the properties of the action map $f$ 
are of considerable interest 
but are hard if not impossible to study in general. In the simplest case,
$\Gamma_0$ is a quadratic functional of the field $\phi$ (reasonably
chosen to ensure that the functional integral is well defined).
Then, the functional integral is Gaussian and one immediately finds 
(free field theory; $const.$ is some constant depending on the
choice of $C$)
\begin{eqnarray}
\label{M4}
\Gamma [\phi] &=& \Gamma_0 [\phi]\ +\ const.\ . 
\end{eqnarray}
There are very few other cases in which the formalism can explicitly be 
studied beyond perturbation theory. A number of exact results exist 
in quantum mechanics (which can be understood as quantum field theory
in 0+1-dimensional space-time; see, e.g., \cite{gros,klei}).
For some quantum field theoretic results see, e.g., \cite{abda}.\\

It is common and successful practice in mathematics and physics to 
approach difficult infinite-dimensional problems from their finite-dimensional
ana\-logues. 
For example, in numerical studies within the framework of lattice 
quantum field theory the infinite-dimensional functional integral 
as present
in eq.\ (\ref{M3}) is replaced by a multidimensional multiple integral.
The simplest finite-dimensional analogue of eq.\ (\ref{M3}) is being obtained
by replacing the infinite-dimensional functional integral by
an one-dimensional integral over the real line (More precisely,
we obtain it from the Euclidean field theory version of eq.\ (\ref{M3})
where the imaginary unit $i$ in the exponent is replaced by (-)1.
$g^\prime$ denotes here
the first derivative of the function $g$.).
\begin{eqnarray}
\label{M5}
{\rm e}^{\displaystyle\ g(y)}\ &=& C\
\int\limits^{+\infty}_{-\infty} dx\ \ {\rm e}^{\displaystyle\ g_0(x+y) 
\ -\ g^\prime (y) x}
\end{eqnarray}
Still, even the study of eq.\ (\ref{M5}) represents a formidable task.
The consideration of the (one-dimensional) analogues of the eqs.\ 
(\ref{M1a})-(\ref{M3}) is often pursued under the name of
zero-dimensional field theory 
\cite{prok1}-\cite{klau3},
\cite{itzy}, Subsect.\ 9-4-1,
p.\ 463,
\cite{zinn1}-\cite{bend3},
\cite{cvit2},
\cite{chal}-\cite{bend2},
\cite{klau2}, Chap.\ 9, p.\ 211, \cite{malb}-\cite{haeu} 
\footnote{We have included into the list of
reference also articles on the static ultralocal single-component
scalar model but left aside papers on the corresponding
$O(N)$ symmetric model.}.\\

For simplicity, 
the above discussion has been based on the consideration of a bosonic
quantum field. However, fermionic (Grassmann valued) quantum fields 
are also of considerable physical interest (for a general discussion
of Grassmann variables see \cite{bere1}). The analogue of 
eq.\ (\ref{M3}) for a purely fermionic field theory of the Grassmann
field $\Psi$, $\bar\Psi$ reads as follows.
\begin{eqnarray}
\label{M6a}
{\rm e}^{\displaystyle\ i \Gamma [\bar\Psi,\Psi ]}\ &=& C\
\int D\left(\chi,\bar\chi\right)\ \ {\rm e}^{\displaystyle\ i\Gamma_0 
[\bar\chi + \bar\Psi,\chi + \Psi ]}\nonumber\\[0.3cm]
&&\hspace{2cm}\times\ {\rm e}^{\displaystyle\ i 
\int d^kx\ \left(\bar\eta(x) \chi(x) + \bar\chi(x) \eta(x) \right)}\\[0.3cm]
\label{M6b}
\bar\eta(x) &=&\frac{\delta\Gamma [\bar\Psi,\Psi ]}{\delta\Psi(x)}\ ,\
\ \ \ \ 
\eta(x)\ =\ -\ \frac{\delta\Gamma [\bar\Psi,\Psi ]}{\delta\bar\Psi(x)}
\end{eqnarray}
Here, $D\left(\chi,\bar\chi\right)$ denotes the infinite-dimensional
Grassmann integration and
the functional derivatives used in (\ref{M6b}) are left
Grassmann derivatives. The finite-dimensional (fermionic) analogues of the 
eqs.\ (\ref{M1a})-(\ref{M2}) and (\ref{M6a}), (\ref{M6b}) consequently 
read\footnote{Again,
we obtain them from the Euclidean field theory version of these equations
where in eq.\ (\ref{M1a}) the imaginary unit $i$ in the exponent 
is replaced by 1 (In the Grassmann algebra case this is just a matter
of convention as convergence considerations for integrals do not
play any role.). $Z[\{\bar\eta\},\{\eta\}]$ is the Fourier-Laplace 
transform of $\exp G_0[\{\bar\chi\},\{\chi\} ]$. For a discussion
of the Fourier-Laplace transformation in a Grassmann algebra see
\cite{love1}, \cite{vasi}, Chap.\ 1, \S 6, Subsect.\ 2, pp.\ 62/63
(Subsect. 1.6.2, p.\ 63 of the English translation),
\cite{bere2}, Appendix A, p.\ 355, \cite{lase}, Appendix A,
p.\ 3709, \cite{roep}, Subsect.\ 10.5.4., p.\ 339,
\cite{khre1}, Subsect.\ 2.3, p.\ 72 (p.\ 74 of the English translation).}
\begin{eqnarray}
\label{M7a0}
Z[\{\bar\eta\},\{\eta\}] &=& C\ 
\int \prod_{l=1}^n\left(d\chi_l\ d\bar\chi_l\right)
\ \ {\rm e}^{\displaystyle\ 
G_0[\{\bar\chi\},\{\chi\} ]\ +\
\sum_{l=1}^n\ \left(\bar\eta_l \chi_l + 
\bar\chi_l \eta_l \right)}\ \ ,\ \ \ \ \ \ \ \\[0.3cm]
\label{M7b0}
W[\{\bar\eta\},\{\eta\}] &=& \ln Z[\{\bar\eta\},\{\eta\}]\ ,\\[0.3cm]
\label{M7c0}
G[\{\bar\Psi\},\{\Psi\} ] &=& W[\{\bar\eta\},\{\eta\}]
- \sum_{l=1}^n\ \left(\bar\eta_l \Psi_l + 
\bar\Psi_l \eta_l \right)\ ,\\[0.3cm]
\label{M7d0}
\bar\Psi_l &=& -\ {\partial W[\{\bar\eta\},\{\eta\}]\over \partial \eta_l}\ ,
\ \ \ \
\Psi_l \ =\ {\partial W[\{\bar\eta\},\{\eta\}]\over \partial \bar\eta_l}
\ ,
\end{eqnarray}
and
\begin{eqnarray}
\label{M7a}
{\rm e}^{\displaystyle\ G[\{\bar\Psi\},\{\Psi\} ]}\ &=& C\
\int \prod_{l=1}^n\left(d\chi_l\ d\bar\chi_l\right)
\ \ {\rm e}^{\displaystyle\ 
G_0[\{\bar\chi + \bar\Psi\},\{\chi + \Psi\} ]}\nonumber\\[0.3cm]
&&\hspace{2cm}\times\ {\rm e}^{\displaystyle\ \
\sum_{l=1}^n\ \left(\bar\eta_l \chi_l + 
\bar\chi_l \eta_l \right)}\ \ ,\\[0.3cm]
\label{M7b}
\bar\eta_l &=&
\frac{\partial G[\{\bar\Psi\},\{\Psi\} ]}{\partial\Psi_l}\ \ ,\ \ \ \
\eta_l\ =\ -\ 
\frac{\partial G[\{\bar\Psi\},\{\Psi\} ]}{\partial\bar\Psi_l}\ \ ,
\end{eqnarray}
respectively.
$\{\bar\Psi\}$, $\{\Psi\}$ denote the sets of Grassmann variables 
$\bar\Psi_l$, $l= 1,\ldots,n$ and $\Psi_l$, $l= 1,\ldots,n$, respectively,
which are the generators of the Grassmann algebra ${\cal G}_{2n}$
(more precisely, we are considering a Grassmann algebra ${\cal G}_{4n}$
as the $\chi_l$, $\bar\chi_l$ in eq.\ (\ref{M7a}) are also
Grassmann variables, but we will ignore this mathematical subtlety
in the following).
These generators obey the standard relations
\begin{eqnarray}
\label{M8}
\Psi_l \Psi_m + \Psi_m \Psi_l&=& \bar\Psi_l \Psi_m + \Psi_m \bar\Psi_l
\ =\ \bar\Psi_l \bar\Psi_m + \bar\Psi_m \bar\Psi_l\ =\ 0\ .
\end{eqnarray}
In this paper, we will concentrate on the explicit study of the eqs.\ 
(\ref{M7a}), (\ref{M7b}) for small values of $n$ ($n=2,3,4)$
(Some of the calculations have been performed by means of a
purpose designed Mathematica programme \cite{wolf}.). 
The eqs.\ (\ref{M7a}), (\ref{M7b})
define (implicitly) a map $f$ between the elements $G_0$ and $G$
of the Grassmann algebra ${\cal G}_{2n}$
(In analogy to the infinite-dimensional 
case, we call the map $f$ the `action map'.). As we will see, 
the eqs.\ (\ref{M7a}), (\ref{M7b}) are equivalent to a coupled
system of nonlinear matrix equations which however can successively
be solved completely (for a general exposition of matrix
equations see, e.g., \cite{horn,maly}). This way, we will explicitly work
out the action map $f$ for the following fairly general Ansatz for $G_0$.
\begin{eqnarray}
\label{M9}
&&\hspace{-2cm}G_0[\{\bar\Psi\},\{\Psi\} ]\nonumber\\[0.3cm]
&=&A^{(0)}\ +\ \sum_{l,m=1}^n A_{l,m}^{(2)} 
\bar\Psi_l \Psi_m\nonumber\\[0.3cm]
&&\ +\ \left(\frac{1}{2!}\right)^2
\sum_{l_1,l_2,m_1,m_2=1}^n A_{l_1 l_2,m_1 m_2}^{(4)} 
\bar\Psi_{l_1}\bar\Psi_{l_2} \Psi_{m_1}\Psi_{m_2}\nonumber\\[0.3cm]
&&\ +\ \left(\frac{1}{3!}\right)^2
\sum_{l_1,l_2,l_3,m_1,m_2,m_3=1}^n A_{l_1 l_2 l_3,m_1 m_2 m_3}^{(6)} 
\bar\Psi_{l_1}\bar\Psi_{l_2}\bar\Psi_{l_3} 
\Psi_{m_1}\Psi_{m_2}\Psi_{m_3}\nonumber\\[0.3cm]
&&\ +\ \ldots\nonumber\\[0.3cm]
&&\ +\ \left(\frac{1}{n!}\right)^2
\sum_{l_1,\ldots,l_n,m_1,\ldots,m_n=1}^n 
A_{l_1 \ldots l_n,m_1 \ldots m_n}^{(2n)} 
\bar\Psi_{l_1}\ldots\bar\Psi_{l_n} 
\Psi_{m_1}\ldots\Psi_{m_n}
\end{eqnarray}
Here, $A^{(0)}$ is some constant and
the coefficients $A_{\ldots}^{(2k)} $, $k>1$, 
are chosen to be completely 
antisymmetric in the first and in the second half of their indices,
respectively.\\

Although the explicit determination of the action map $f$ between
$G_0$ and $G$ for low-dimensional Grassmann algebras
represents previously unknown information, it may seem that the study
of the map $f$ for low-dimensional Grassmann algebras is a 
mathematical exercise of purely academic nature as quantum field theory and
statistical mechanics are concerned with infinitely many degrees
of freedom. To some extent this view may be justified 
for the time being but one should 
also take note of the fact that results for the Grassmann algebras
${\cal G}_{2n}$ and ${\cal G}_{2(n-1)}$ are closely related. To see this
observe the following. Put in eq.\ (\ref{M9}) considered in the case
of the Grassmann algebra ${\cal G}_{2n}$ the coefficient 
$A_{n,n}^{(2)}$ equal to one
but all other coefficients $A_{\ldots}^{(2k)} $, $k>1$, equal to zero
whose index set $\{\ldots\} $ contains at least one index with value $n$.
\begin{eqnarray}
\label{M9a1}
A_{n,n}^{(2)}&=&1\\[0.3cm]
\label{M9a2}
A_{\ldots n\ldots}^{(2k)}&=&0,\ k > 1
\end{eqnarray}
Then, perform in eq.\ (\ref{M7a}) the Grassmann integrations with
respect to $\chi_n$, $\bar\chi_n$. Up to the factor 
$(\exp\bar\Psi_n\Psi_n)$ present on both sides 
(no summation with respect to $n$ here) eq.\ (\ref{M9})
then coincides with eq.\ (\ref{M9}) considered in the case of the
Grassmann algebra ${\cal G}_{2(n-1)}$. Consequently, results obtained
for low-dimensional Grassmann algebras tightly constrain 
structures to be found for Grassmann algebras ${\cal G}_{2n}$
with arbitrarily chosen $n$. In fact, we will use this observation
in two ways. On hand side, we will rely on it in order to check
the explicit results obtained for $n=4$ and $n=3$ for compatibility
with those obtained for $n=3$ and $n=2$, respectively.
On the other hand, on the basis of the above observation we will 
extrapolate some results obtained for $n=2,3,4$ to arbitrary $n$
which can be used later in the future as working hypothesis for further
studies.\\

Having explicitly worked out the action map $f$ between
$G_0$ and $G$ for low-dimensional Grassmann algebras, we will
not stop our investigation at this point 
but pursue our study still one step further.
In \cite{prok1,scha1}\footnote{Only by accident, in preparing the
final draft of the present paper we became
aware of the fact that the concept proposed in \cite{scha1}
(preprint version: University of Leipzig Preprint NTZ 16/1993,
Physics E-Print Archive: hep-th/9312137) has also been proposed
earlier by Prokhorov \cite{prok1}.} it has been argued
(in a quantum field theoretic context), 
that it might be physically
sensible and interesting to look for actions $\Gamma_0[\phi]$ which 
are not quadratic functionals of the field $\phi$ (i.e., which
do not describe free fields) but for which eq.\ 
(\ref{M4}) also applies. For the purpose of the present investigation
we will slightly extend our search. We will look for solutions to the 
equation ($0<\lambda\in{\bf R}$)
\begin{eqnarray}
\label{M9b}
G[\{\bar\Psi\},\{\Psi\} ]&=&G_0[\{\lambda\bar\Psi\},\{\lambda\Psi\} ]
\ +\ \Delta_f(\lambda)\ .
\end{eqnarray}
$\lambda$ can be considered here as a finite-dimensional analogue
of a wave function renormalization constant in continuous space-time
quantum field theory. $\Delta_f(\lambda)$
is some constant which is allowed to depend on $\lambda$.
Eq.\ (\ref{M9b}) turns the implicit
representation of the map $f$ given by the eqs.\ (\ref{M7a}),
(\ref{M7b}) into a Grassmann integral equation for 
$G_0[\{\bar\Psi\},\{\Psi\} ]$ (more precisely, into a
nonlinear Grassmann integro-differential equation). 
As we will see, this Grassmann
integral equations is equivalent to a coupled system of nonlinear
matrix equations whose solution in turn is equivalent to the solution
of the considered Grassmann integral equation. 
In the present study, to us eq.\ (\ref{M9b}) is just a mathematical 
problem to be studied. The possible relevance of any solution of eq.\ 
(\ref{M9b}) to physical problems will remain beyond the scope
of the present paper. Some comments in this respect can 
be found in \cite{prok1,scha1}.\\

The plan of the paper is as follows. In Sect.\ 2 we work out
explicitly the action map $f$ between $G_0$ and $G$. Subsect.\ 2.1
contains some mathematical preliminaries while the following three
subsections are devoted to the cases $n=2,3,4$, respectively.
Subsect.\ 2.5 finally studies the extrapolation of some
of the results obtained to Grassmann algebras ${\cal G}_{2n}$
with arbitrarily chosen $n$.
Sect.\ 3 is concerned with the study of the Grassmann integral 
equation (\ref{M9b}). On the basis of the results obtained 
in Sect.\ 2, in Subsects.\ 3.1-3.3 it is solved for 
$n=2,3,4$, respectively. Then, Subsect.\ 3.4 contains an analysis of 
certain aspects of the solutions of the Grassmann integral
equation found for $n=4$.
In Sect.\ 4 the discussion of the results 
and conclusions can be found. The paper is supplemented 
by three Appendices.\\

\section{The action map for low-dimensional Grass-\hfill\break
mann algebras}

\subsection{Some definitions}

To simplify the further considerations we introduce a set of
${n\choose k}\times {n\choose k}$ matrices ${\sf A}^{(2k)}$ ($k = 1,\ldots,n$) 
by writing (choose $l_1< l_2< \ldots <l_k$, $m_1< m_2< \ldots <m_k$)
\begin{eqnarray}
\label{M10a}
{\sf A}^{(2k)}_{LM} &=& A_{l_1 \ldots l_k,m_1 \ldots m_k}^{(2k)}\ 
\end{eqnarray}
(We identify the indices $L$, $M$ with the ordered strings
$l_1 \ldots l_k$, $m_1 \ldots m_k$.)
or, more generally (not requesting 
$l_1< l_2< \ldots <l_k$, $m_1< m_2< \ldots <m_k$)
\begin{eqnarray}
\label{M10b}
{\sf A}^{(2k)}_{LM}\ =\ {\rm sgn}\left[\sigma_a(l_1, \ldots, l_k)\right]\
{\rm sgn}\left[\sigma_b(m_1, \ldots, m_k)\right]\
A_{l_1 \ldots l_k,m_1 \ldots m_k}^{(2k)}\ .
\end{eqnarray}
The indices $L,M$ label the equivalence classes of all permutations of the 
indices $l_1, \ldots, l_k$ and $m_1, \ldots, m_k$, respectively, and $\sigma_a$, $\sigma_b$
are the permutations which bring the indices $l_i, m_i$ ($i = 1,\ldots,k$)
into order with respect to the $<$ relation
(i.e., $\sigma_a(l_1)< \sigma_a(l_2)< \ldots <\sigma_a(l_k)$, 
$\sigma_b(m_1)< \sigma_b(m_2)< \ldots <\sigma_b(m_k)$). 
The matrix elements of the matrix ${\sf A}^{(2k)}$ are arranged 
according to the lexicographical order of the row and column indices 
$L$, $M$
(We identify the indices $L$, $M$ with the ordered strings
$\sigma_a(l_1) \ldots \sigma_a(l_k)$, $\sigma_b(m_1) \ldots \sigma_b(m_k)$,
respectively.).\\

We also define a set of (dual)
${n\choose k}\times {n\choose k}$ matrices 
${\sf A}^{(2k) \star}$ ($k = 1,\ldots,n$) by writing
\begin{eqnarray}
\label{M11}
{\sf A}^{(2k) \star} &=&{\cal E}^{(k)}  {\sf A}^{(2k) T} {\cal E}^{(k) T}
\end{eqnarray}
where the ${n\choose k}\times {n\choose k}$ matrix ${\cal E}^{(k)}$
is defined by 
\begin{eqnarray}
\label{M12a}
{\cal E}^{(k)}_{LM}&=&\epsilon_{l_1 \ldots l_{n-k} m_1 \ldots m_k}\ ,
\end{eqnarray}
consequently,
\begin{eqnarray}
\label{M12b}
{\cal E}^{(k) T}&=&(-1)^{(n-k)k}\ {\cal E}^{(n-k)}
\end{eqnarray}
(Quite generally, for any ${n\choose k}\times {n\choose k}$ matrix ${\sf B}$
we define ${\sf B}^\star$ by ${\sf B}^\star = 
{\cal E}^{(k)}  {\sf B}^T {\cal E}^{(k) T}$.).
It holds (${\bf 1}_r$ is the $r\times r$ unit matrix)
\begin{eqnarray}
\label{M13a}
{\cal E}^{(k)}\ {\cal E}^{(k) T}&=&{\bf 1}_{n\choose k}\ ,\\[0.3cm]
\label{M13b}
{\cal E}^{(k) T}\ {\cal E}^{(k)}&=&{\bf 1}_{n\choose k}\ .
\end{eqnarray}
The transition from a matrix ${\sf B}$ to the matrix ${\sf B}^\star$ 
corresponds to applying the Hodge star operation to the two subspaces
of the Grassmann algebra ${\cal G}_{2n}$ generated by the two
sets of Grassmann variables $\{\bar\Psi\}$ and $\{\Psi\}$ and
interchanging them
(cf., e.g., \cite{marc1}, Part II, Chap.\ 4, p.\ 50).
This operation on the matrix ${\sf B}$ is an involution as
$\left({\sf B}^\star\right)^\star = {\sf B}$.\\

Furthermore, it turns out to be convenient to define arrays of partition
functions (i.e., their finite-dimensional analogues).
First, we choose 
\begin{eqnarray}
\label{M14a}
C&=&{\rm e}^{\displaystyle - G_0[\{0\},\{0\} ]}\ =\ 
{\rm e}^{\displaystyle - A^{(0)}}\ .
\end{eqnarray}
This choice in effect cancels any constant term in eq.\ (\ref{M9})
(in this respect also see \cite{scha1}, p.\ 288).
Now, we define (We apply the convention 
$\int d\chi_i\ \chi_j = \delta_{ij}$.)\footnote{The 
notation ${\sf P}^{(2n) \star}$
is chosen with hindsight.
Of course, ${\sf P}^{(2n) \star}\ =\ {\sf P}^{(2n)}$
--- ignoring the fact that (very formally) these constants live in different
spaces, cf.\ eq.\ (\ref{M11}): ${\sf P}^{(2n) \star}$ is a constant
without indices while ${\sf P}^{(2n)}$ is a $1\times 1$
matrix with a length $n$ row and column index.}
\begin{eqnarray}
\label{M14b}
P&=&{\sf P}^{(2n) \star}\ =\
C\ \int \prod_{l=1}^n\left(d\chi_l\ d\bar\chi_l\right)
\ \ {\rm e}^{\displaystyle\ 
G_0[\{\bar\chi\},\{\chi\} ]}\ .
\end{eqnarray}
We then define arrays of partition functions ${\sf P}^{(2n-2k) \star}$
(These are ${n\choose k}\times{n\choose k}$ matrices.)
for subsystems of Grassmann variables
where $k$ degrees of freedom have been omitted (In slight misuse of physics
terminology  we denote a pair of Grassmann variables $\bar\Psi_l$, $\Psi_m$
by the term `degree of freedom'. $l_1 < l_2 <\ldots < l_k$, 
$m_1 < m_2 <\ldots < m_k$ in the following).
\begin{eqnarray}
\label{M15a}
{\sf P}^{(2n-2k) \star}_{LM}&=&\frac{\partial}{
\partial A_{l_1,m_1}^{(2)}}\ldots
\frac{\partial}{\partial A_{l_k,m_k}^{(2)}}\ P\\[0.3cm]
\label{M15b}
&=&(-1)^k\ \frac{\partial}{\partial\eta_{l_1}}\
\frac{\partial}{\partial\bar\eta_{m_1}}\ldots \frac{\partial}{\partial\eta_{l_k}} 
\ \frac{\partial}{\partial\bar\eta_{m_k}}\
Z[\{\bar\eta\},\{\eta\} ]\Bigg|_{\bar\eta\ =\ \eta\ =\ 0}
\end{eqnarray}
Recursively, eq.\ (\ref{M15a}) can be written as follows
($l_k > l_{k-1}$, $m_k > m_{k-1}$;
note the different meaning of the indices $L$, $M$ on the l.h.s.\ and 
on the r.h.s.\ of the equation.).
\begin{eqnarray}
\label{M17}
{\sf P}^{(2n-2k) \star}_{LM}&=&
\frac{\partial\ 
{\sf P}^{(2n-2k+2) \star}_{LM}}{\partial A_{l_k,m_k}^{(2)}}
\end{eqnarray}

Let us illustrate the above definitions by means of a simple example.
Choose
\begin{eqnarray}
\label{M18}
G_0[\{\bar\chi\},\{\chi\} ]
&=&\sum_{l,m=1}^n A_{l,m}^{(2)} \bar\chi_l \chi_m\ .
\end{eqnarray}
Then
\begin{eqnarray}
\label{M19}
Z[\{\bar\eta\},\{\eta\} ]&=&\det{\sf A}^{(2)}\ 
{\rm e}^{\displaystyle -\ \bar\eta\left[{\sf A}^{(2)}\right]^{-1}\eta}
\end{eqnarray}
and
\begin{eqnarray}
\label{M20}
{\sf P}^{(2n-2k) \star}&=&C^{\; n-k}\left({\sf A}^{(2)}\right)
\end{eqnarray}
(cf.\ the references cited in relation to eq.\ (\ref{A1f}) of
Appendix A and \cite{kerl1}, Sect.\ 2, \cite{char1},
also see \cite{zinn2}, Chap.\ 1, Sect.\ 1.9).
Here, $C^{\; n-k}\left({\sf A}^{(2)}\right)$ is the $(n-k)$th 
supplementary compound matrix
of the matrix ${\sf A}^{(2)}$ (for a definition and some properties
of compound matrices see Appendix A).
By virtue of eq.\ (\ref{A2}) (see Appendix A) it holds
\begin{eqnarray}
\label{M21}
{\sf P}^{(2n-2k) \star} {\sf P}^{(2k)}\ =\
{\sf P}^{(2k)} {\sf P}^{(2n-2k) \star}
&=&\det {\sf A}^{(2)}\ {\bf 1}_{n\choose k}\ .
\end{eqnarray}

\subsection{Explicit calculation: $n=2$}

The case of the Grassmann algebra ${\cal G}_4$ ($n=2$) to be treated 
in the present subsection is still algebraically fairly simple but 
already exhibits many of the features which we will meet in considering
the larger Grassmann algebras ${\cal G}_6$, ${\cal G}_8$. 
Therefore, to some extent this subsection serves a didactical
purpose in order to give the reader a precise idea of the calculations
to be performed in the following two subsections. These calculations
will proceed exactly by the same steps as in this subsection
but the algebraic 
complexity of the expressions will grow considerably. Also from
a practical, calculational point of view it is advisable to choose
an approach which proceeds stepwise from the most simple case ($n=2$) to
the more involved ones ($n=3, 4$) in order to accumulate experience
in dealing with this growing complexity.
On the other hand, the case $n=2$ is special in some respect
and deserves attention in its own right.\\

According to our general Ansatz (\ref{M9}) we put
\begin{eqnarray}
\label{NB1}
G_0[\{\bar\Psi\},\{\Psi\} ]
&=&A^{(0)}\ +\ \sum_{l,m=1}^2 A_{l,m}^{(2)} \bar\Psi_l \Psi_m\ +\ 
A_{1 2,1 2}^{(4)} \bar\Psi_1\bar\Psi_2 \Psi_1\Psi_2
\end{eqnarray}
and $G[\{\bar\Psi\},\{\Psi\} ]$ can be written in the same way. 
\begin{eqnarray}
\label{NB2}
G[\{\bar\Psi\},\{\Psi\} ]
&=&A^{(0) \prime}\ +\ \sum_{l,m=1}^2 A_{l,m}^{(2) \prime} 
\bar\Psi_l \Psi_m\ +\ 
A_{1 2,1 2}^{(4) \prime} 
\bar\Psi_1\bar\Psi_2 \Psi_1\Psi_2
\end{eqnarray}
No other terms will occur for symmetry reasons.
One quickly finds for the partition function
(cf.\ eq.\ (\ref{M14b}))
\begin{eqnarray}
\label{NB3}
P\ =\ {\rm e}^{\displaystyle\ A^{(0) \prime}}\ =\ 
{\sf P}^{(4) \star}&=&\det{\sf A}^{(2)}\ -\ 
{\sf A}^{(4) \star}\ .
\end{eqnarray}
Of course, here
${\sf A}^{(4) \star} = A_{1 2,1 2}^{(4)}$ applies ---
again ignoring the fact that (very formally) these constants live in different
spaces, cf.\ eq.\ (\ref{M11}).
The notation ${\sf P}^{(4) \star}$ is introduced in order to indicate
how in larger Grassmann algebras this partition function transforms
under linear (unitary) transformations of the two subsets $\{\bar\Psi\}$, 
$\{\Psi\}$ of the generators of the Grassmann
algebra. Clearly, ${\sf P}^{(4) \star}$ then
transforms exactly the same way as ${\sf A}^{(4) \star}$ does and this fact
suggests the chosen notation (The same will apply to any other partition 
function ${\sf P}^{(2n) \star}$ for larger Grassmann algebras 
${\cal G}_{2n}$.). The result of the map $g_2\circ g_1$
reads
\begin{eqnarray}
\label{NB4}
W[\{\bar\eta\},\{\eta\}] &=& \ln {\sf P}^{(4) \star}\ -\
\sum_{l,m=1}^2 \frac{\left({\rm adj}\; {\sf A}^{(2)}\right)_{lm}}{
{\sf P}^{(4) \star}}\ \bar\eta_l \eta_m\ +
\ \frac{A_{1 2,1 2}^{(4)}}{
\left({\sf P}^{(4) \star}\right)^2}\ \bar\eta_1\bar\eta_2 \eta_1\eta_2\ .
\end{eqnarray}
The only assumption made to arrive at this result is that 
${\sf P}^{(4) \star}\not= 0$.
We can now proceed on the basis of the general eq.\ (\ref{M7c0})
specified to $n = 2$.
\begin{eqnarray}
\label{NB5}
G[\{\bar\Psi\},\{\Psi\}]&=&
W[\{\bar\eta\},\{\eta\}]\ -
\sum_{l=1}^2\ \left(\bar\eta_l \Psi_l + \bar\Psi_l \eta_l \right)
\end{eqnarray}
We insert eq.\ (\ref{NB2}) onto the l.h.s.\ of eq.\ (\ref{NB5})
and the explicit expressions for $\bar\eta$, $\eta$ found from eq.\
(\ref{NB2}) according to eq.\ (\ref{M7b}) on its r.h.s.. Comparing
coefficients on both sides we find the following two coupled equations.
\begin{eqnarray}
\label{NB6a}
{\sf A}^{(2) \prime} &=& 2 {\sf A}^{(2) \prime}\ -\ {\sf A}^{(2) \prime}\
\frac{{\rm adj}\; {\sf A}^{(2)} }{{\sf P}^{(4) \star}}\ {\sf A}^{(2) \prime}
\\[0.3cm]
\label{NB6b}
A_{1 2,1 2}^{(4) \prime} &=& 4 A_{1 2,1 2}^{(4) \prime}\ -\ 2\
\frac{{\rm tr}\left[{\sf A}^{(2) \prime}\ 
{\rm adj}\; {\sf A}^{(2)} \right]}{{\sf P}^{(4) \star}}
\ A_{1 2,1 2}^{(4) \prime}\ +\ 
\left(\frac{\det{\sf A}^{(2) \prime}}{{\sf P}^{(4) \star}}\right)^2\ 
A_{1 2,1 2}^{(4)}
\end{eqnarray}
Eq.\ (\ref{NB6a}) can immediately be simplified to read
\begin{eqnarray}
\label{NB7}
{\sf A}^{(2) \prime} &=&{\sf A}^{(2) \prime}\
\frac{{\rm adj}\; {\sf A}^{(2)} }{{\sf P}^{(4) \star}}\ 
{\sf A}^{(2) \prime}\ .
\end{eqnarray}
From eq.\ (\ref{NB7}) one recognizes that the matrix 
${\sf A}^{(2) \prime}$ is the generalized \{2\}-inverse of the matrix
${\rm adj}\; {\sf A}^{(2)}/{\sf P}^{(4) \star}$ (cf., e.g.,
\cite{beni}, Chap.\ 1, p.\ 7).\\

We can now successively solve the eqs.\ (\ref{NB6a}), (\ref{NB6b}).
Choosing $\det{\sf A}^{(2) \prime}\not=0$ (By virtue of eq.\ (\ref{NB7})
this entails
$\det{\sf A}^{(2)}\not=0$.), we immediately find from eq.\
(\ref{NB7})
\begin{eqnarray}
\label{NB8}
{\sf A}^{(2) \prime} &=& \left(\frac{{\sf P}^{(4) \star}}{
\det{\sf A}^{(2)}}\right)\  {\sf A}^{(2)}\ .
\end{eqnarray}
Inserting this expression for ${\sf A}^{(2) \prime}$ into eq.\ (\ref{NB6b})
yields the following solution.
\begin{eqnarray}
\label{NB9}
A_{1 2,1 2}^{(4) \prime}&=&\left(\frac{{\sf P}^{(4) \star}}{
\det{\sf A}^{(2)}}\right)^2\ A_{1 2,1 2}^{(4)}
\end{eqnarray}
In analogy to eq.\ (\ref{NB3}), we can now define a quantity 
\begin{eqnarray}
\label{NB10}
{\sf P}^{(4) \star\prime}&=&\det{\sf A}^{(2) \prime}\ -\ 
{\sf A}^{(4) \star\prime}
\end{eqnarray}
and from eqs.\ (\ref{NB8}), (\ref{NB9}) we find (taking into
account eq.\ (\ref{NB3}))
\begin{eqnarray}
\label{NB11}
{\sf P}^{(4) \star\prime}
&=&\frac{\left({\sf P}^{(4) \star}\right)^3}{
\left(\det{\sf A}^{(2)}\right)^2}
\ =\ \left(\frac{{\sf P}^{(4) \star}}{\det{\sf A}^{(2)}}\right)^2\ 
{\sf P}^{(4) \star}\ .
\end{eqnarray}
Taking the determinant on both sides of eq.\ (\ref{NB8}) provides us
with the following useful relation.
\begin{eqnarray}
\label{NB12}
\det{\sf A}^{(2) \prime} &=&
\frac{\left({\sf P}^{(4) \star}\right)^2}{\det{\sf A}^{(2)}}
\end{eqnarray}
Up to this point, very little is special to the case $n=2$
and we will meet the analogous equations in the next subsections.\\

We turn now to some features which are closely related to the 
algebraic simplicity of the case $n=2$ and which cannot easily
be identified in larger Grassmann algebras.
The eqs.\ (\ref{NB11}) and (\ref{NB12}) can now be combined to
yield the equation
\begin{eqnarray}
\label{NB13}
{\sf P}^{(4) \star\prime}
&=&\frac{\det{\sf A}^{(2) \prime}}{\det{\sf A}^{(2)}}\ 
{\sf P}^{(4) \star}
\end{eqnarray}
which is converted (${\sf P}^{(4) \star}, \det{\sf A}^{(2) \prime}\not=0$ entail
${\sf P}^{(4) \star\prime} \not=0$.) into
\begin{eqnarray}
\label{NB14}
\frac{\det{\sf A}^{(2) \prime}}{{\sf P}^{(4) \star\prime}}&=&
\frac{\det{\sf A}^{(2)}}{{\sf P}^{(4) \star}}\ .
\end{eqnarray}
An equivalent form of eq.\ (\ref{NB14}) is
\begin{eqnarray}
\label{NB15}
\frac{A_{1 2,1 2}^{(4) \prime}}{\det{\sf A}^{(2) \prime} }\ =\ 
\frac{A_{1 2,1 2}^{(4)}}{\det{\sf A}^{(2)}}\ .
\end{eqnarray}
From eqs.\ (\ref{NB14}), (\ref{NB15}) we recognize that for $n=2$
the action map $f$ has an invariant which can be calculated 
from the left or right hand sides of these equations.\\

We are now going to invert the action map $f$\ \footnote{
\label{footn3}Speaking rigorously and for any value of $n$, 
this is not possible as due to our choice (\ref{M14a})
$A^{(0)}$ cannot be found from $G[\{\bar\Psi\},\{\Psi\} ]$. 
The action map $f$ can only be inverted in the subspace of the 
coefficients $A^{(2k) \prime}$, $A^{(2k)}$, $k>0$.
In a way, $A^{(0)}$ is a dummy variable while 
$A^{(0) \prime} = \ln P$
can be understood as a function of $A^{(2k) \prime}$, $k>0$,
via its dependence on $A^{(2k)}$, $k>0$. In this view,
the coefficients $A^{(2k)}$, $k>0$, are given in terms
of the coefficients $A^{(2k) \prime}$, $k>0$, by means of the
inverse action map $f^{-1}$. On the other hand, $A^{(0)}$
can be given a sensible meaning if one assumes that the action
$G_0$ has been induced by some other action $G_{-1}$ exactly
the same way as the action $G$ is being induced by $G_0$.
Then, one can derive 
an expression for $A^{(0)}$ in terms of the coefficients
$A^{(2k)}$, $k>0$, by means of the inverse action map $f^{-1}$. This way, on the basis of eq.\ (\ref{NB16})
one finds for $n=2$:
$A^{(0)} = 2 \ln\det{\sf A}^{(2)} - \ln{\sf P}^{(4) \star}$
(cf.\ Subsect.\ 3.1, eq.\ (\ref{QB3})).}.
From eqs.\ (\ref{NB11}) and (\ref{NB14}) we easily find 
\begin{eqnarray}
\label{NB16}
{\sf P}^{(4) \star} &=&
\frac{\left(\det{\sf A}^{(2) \prime}\right)^2}{{\sf P}^{(4) \star\prime}}
\ =\ \left(\frac{\det{\sf A}^{(2) \prime}}{
{\sf P}^{(4) \star\prime}}\right)^2\ {\sf P}^{(4) \star\prime}\ .
\end{eqnarray}
And eq.\ (\ref{NB14}) also allows us to find the following inversion
formulas for the map $f$ from eqs.\ (\ref{NB8}), (\ref{NB9}).
\begin{eqnarray}
\label{NB17a}
{\sf A}^{(2)} &=& 
\left(\frac{\det{\sf A}^{(2) \prime} }{{\sf P}^{(4) \star\prime}}\right)\  
{\sf A}^{(2) \prime}\\[0.3cm]
\label{NB17b}
A_{1 2,1 2}^{(4)} 
&=&\left(\frac{\det{\sf A}^{(2) \prime}}{{\sf P}^{(4) \star\prime}}\right)^2\ 
A_{1 2,1 2}^{(4) \prime} 
\end{eqnarray}
From the above equations we see that for $n=2$ the action map
$f$ can easily be inverted (once one assumes ${\sf P}^{(4) \star}\not=0$,
$\det{\sf A}^{(2)}\not=0$, ${\sf P}^{(4) \star\prime}\not=0$,
$\det{\sf A}^{(2) \prime}\not=0$).\\

\subsection{Explicit calculation: $n=3$}

The case $n=3$ is already considerably more involved in comparison
with the case $n=2$ treated in the previous subsection. In the 
present and the next subsections, as far as possible and appropriate
we will apply the same wording as in Subsect.\ 2.2 in order
to emphasize their close relation.\\

We start by parametrizing $G_0$ and $G$ according to our
general Ansatz (cf.\
eq.\ (\ref{M9}) and eqs.\ (\ref{NB1}), (\ref{NB2})).
\begin{eqnarray}
\label{NC1}
G_0[\{\bar\Psi\},\{\Psi\} ]
&=&A^{(0)}\ +\ \sum_{l,m=1}^3 A_{l,m}^{(2)} 
\bar\Psi_l \Psi_m\nonumber\\[0.3cm]
&&\ +\ \frac{1}{4}
\sum_{l_1,l_2,m_1,m_2=1}^3 A_{l_1 l_2,m_1 m_2}^{(4)} 
\bar\Psi_{l_1}\bar\Psi_{l_2} \Psi_{m_1}\Psi_{m_2}\ \ \ \ \ \ \ \
\nonumber\\[0.3cm]
&&\ +\ A_{1 2 3,1 2 3}^{(6)} 
\bar\Psi_1\bar\Psi_2\bar\Psi_3 \Psi_1\Psi_2\Psi_3\\[0.3cm]
\label{NC2}
G[\{\bar\Psi\},\{\Psi\} ]
&=&A^{(0) \prime}
\ +\ \sum_{l,m=1}^3 A_{l,m}^{(2) \prime} \bar\Psi_l \Psi_m\nonumber\\[0.3cm]
&&\ +\ \frac{1}{4}
\sum_{l_1,l_2,m_1,m_2=1}^3 A_{l_1 l_2,m_1 m_2}^{(4) \prime} 
\bar\Psi_{l_1}\bar\Psi_{l_2} \Psi_{m_1}\Psi_{m_2}\ \ \ \ \ \ \ \
\nonumber\\[0.3cm]
&&\ +\ A_{1 2 3,1 2 3}^{(6) \prime} 
\bar\Psi_1\bar\Psi_2\bar\Psi_3 \Psi_1\Psi_2\Psi_3
\end{eqnarray}
For the partition function we find
(cf.\ eq.\ (\ref{M14b}))
\begin{eqnarray}
\label{NC4}
P\ =\ {\rm e}^{\displaystyle\ A^{(0) \prime}}\ =\ 
{\sf P}^{(6) \star}&=&\det{\sf A}^{(2)}\ -\ 
{\rm tr}\left({\sf A}^{(4) \star} {\sf A}^{(2)}\right)
-\ {\sf A}^{(6) \star}\\[0.3cm]
\label{NC4b}
&=&- 2 \det{\sf A}^{(2)}\ +\ 
{\rm tr}\left({\sf P}^{(4) \star} {\sf A}^{(2)}\right)
-\ {\sf A}^{(6) \star}\ \ \ \ .\ \ \ \ \ \ \ \ \
\end{eqnarray}
In analogy to eq.\ (\ref{NB3}), 
here ${\sf A}^{(6) \star} = A_{1 2 3,1 2 3}^{(6)}$ applies.
In the lower line (eq.\ (\ref{NC4b})), 
we use the notation (cf.\ eqs.\ (\ref{M15a}) and (\ref{NB3}))
\begin{eqnarray}
\label{NC3c}
{\sf P}^{(4)}&=&C_2\left({\sf A}^{(2)}\right)\ -\ {\sf A}^{(4)},\ \ 
{\sf P}^{(4) \star}\ =\ {\rm adj}\; {\sf A}^{(2)}\ -\ {\sf A}^{(4) \star}
\ .\ \ 
\end{eqnarray}
$\left({\rm adj}\; {\sf A}^{(2)}
\ =\ C_2\left({\sf A}^{(2)}\right)^\star\ \right)$.

After some calculation we obtain the following expression for
$W[\{\bar\eta\},\{\eta\}]$ (To arrive at it we only assume
${\sf P}^{(6) \star}\not= 0$.).
\begin{eqnarray}
\label{NC5}
W[\{\bar\eta\},\{\eta\}] &=& \ln {\sf P}^{(6) \star}\ -\
\frac{{\sf P}_{lm}^{(4) \star}}{{\sf P}^{(6) \star}}\
\bar\eta_l \eta_m\nonumber\\[0.3cm] 
&&-\ \frac{{\sf A}^{(2) \star}_{ML}}{{\sf P}^{(6) \star}}\ 
\bar\eta_{l_1}\bar\eta_{l_2}\eta_{m_1}\eta_{m_2}\
-\ \frac{1}{2}\
\left(\frac{{\sf P}_{lm}^{(4) \star}}{
{\sf P}^{(6) \star}}\ \bar\eta_l \eta_m\right)^2\nonumber\\[0.3cm]
&&+\ \frac{1}{{\sf P}^{(6) \star}}\ 
\left[1\ -\ 
\frac{{\rm tr}\left({\sf P}^{(4) \star} {\sf A}^{(2)} \right)}{
{\sf P}^{(6) \star}}\
+\ \frac{2\det {\sf P}^{(4) \star}}{
\left({\sf P}^{(6) \star}\right)^2}\right]\ 
\bar\eta_1\bar\eta_2\bar\eta_3 \eta_1\eta_2\eta_3\ \ \ \ \ \
\end{eqnarray}
Here and in the following we use
the notation $B_{ML} \bar\eta_{l_1}\bar\eta_{l_2}\eta_{m_1}\eta_{m_2}$
for a multiple sum over $l_1$, $l_2$, $m_1$, $m_2$ with
the restrictions $l_1<l_2$, $m_1<m_2$ applied; 
$L=\{l_1,l_2\}$, $M=\{m_1,m_2\}$.
The analogous convention is also applied to multiple sums over more indices.
To arrive at the further results it is useful to take note
of the equation
\begin{eqnarray}
\label{NC5b}
\left({\sf P}_{lm}^{(4) \star} \bar\eta_l \eta_m\right)^2&=&
-2\ C_2\left({\sf P}^{(4) \star}\right)_{LM}\ 
\bar\eta_{l_1}\bar\eta_{l_2}\eta_{m_1}\eta_{m_2}\ .
\end{eqnarray}
We proceed now exactly the same way as in Subsect.\ 2.2.
We insert eq.\ (\ref{NC2}) onto the l.h.s.\ of eq.\ (\ref{M7c0})
and the explicit expressions for $\bar\eta$, $\eta$ found from eq.\
(\ref{NC2}) according to eq.\ (\ref{M7b}) on its r.h.s.. Again, comparing
coefficients on both sides we find the following three coupled 
nonlinear matrix equations.
\begin{eqnarray}
\label{NC6}
{\sf A}^{(2) \prime} &=& 2 {\sf A}^{(2) \prime}\ -\ 
{\sf A}^{(2) \prime}\
\frac{{\sf P}^{(4) \star} }{{\sf P}^{(6) \star}}\ 
{\sf A}^{(2) \prime}\\[0.3cm]
\label{NC7}
{\sf A}^{(4) \star\prime} &=&
4 {\sf A}^{(4) \star\prime}\ +\ {\sf A}^{(4) \star\prime}\
\frac{{\sf A}^{(2)\prime} {\sf P}^{(4) \star} - 
{\rm tr}\left({\sf A}^{(2)\prime} {\sf P}^{(4) \star}\right) {\bf 1}_3
}{{\sf P}^{(6) \star}}\nonumber\\[0.3cm]
&&+\ \frac{{\sf P}^{(4) \star} {\sf A}^{(2)\prime} -
{\rm tr}\left({\sf P}^{(4) \star} {\sf A}^{(2)\prime}\right) {\bf 1}_3
}{{\sf P}^{(6) \star}}\ {\sf A}^{(4) \star \prime}\nonumber\\[0.3cm]
&&-\ 
{\rm adj}\; {\sf A}^{(2)\prime}\ 
\frac{{\sf P}^{(6) \star} \ {\sf A}^{(2)} - 
{\rm adj}\; {\sf P}^{(4) \star}}{\left({\sf P}^{(6) \star}\right)^2}\ 
{\rm adj}\; {\sf A}^{(2)\prime}\\[0.3cm]
\label{NC8}
A_{1 2 3,1 2 3}^{(6) \prime} &=&
6 A_{1 2 3,1 2 3}^{(6) \prime}\ +\ 
\frac{2}{{\sf P}^{(6) \star}}\Bigg\{ - A_{1 2 3,1 2 3}^{(6) \prime}\
{\rm tr}\left( {\sf P}^{(4) \star} {\sf A}^{(2) \prime} \right)
\nonumber\\[0.3cm]
&&\ +\ 
{\rm tr}\left( {\sf P}^{(4) \star} {\rm adj}\; {\sf A}^{(4) \star\prime}\right)
\ +\ {\rm tr}\left( {\sf A}^{(2)} {\rm adj}\; {\sf A}^{(2) \prime}\right)\ 
{\rm tr}\left( {\sf A}^{(2) \prime} {\sf A}^{(4) \star\prime}\right)
 \nonumber\\[0.3cm]
&&\ -\ \det{\sf A}^{(2) \prime}\ 
{\rm tr}\left( {\sf A}^{(2)} {\sf A}^{(4) \star\prime}\right)
\ +\ \frac{\left(\det {\sf A}^{(2) \prime}\right)^2}{2}
\Bigg\}\nonumber\\[0.3cm]
&&\ +\ \frac{2}{\left({\sf P}^{(6) \star}\right)^2}
\Bigg\{ \det{\sf A}^{(2) \prime}\ 
{\rm tr}\left( {\sf A}^{(4) \star\prime} {\rm adj}\; {\sf P}^{(4) \star}\right)
\nonumber\\[0.3cm]
&&\ -\ {\rm tr}\left[{\rm adj}\left({\sf A}^{(2) \prime} 
{\sf P}^{(4) \star}\right)\right]\ 
{\rm tr}\; \left( {\sf A}^{(2) \prime} {\sf A}^{(4) \star\prime}\right)
\nonumber\\[0.3cm]
&&\ -\ \frac{1}{2} \left(\det {\sf A}^{(2) \prime}\right)^2 
{\rm tr}\left( {\sf P}^{(4) \star} {\sf A}^{(2)}  \right)\Bigg\}
\ +\ \frac{2}{\left({\sf P}^{(6) \star}\right)^3}
\left( \det {\sf A}^{(2) \prime}\right)^2 \det {\sf P}^{(4) \star}\ \ \ \ \ 
\end{eqnarray}
Eq.\ (\ref{NC6}) is equivalent to the equation
\begin{eqnarray}
\label{NC10}
{\sf A}^{(2) \prime} &=&{\sf A}^{(2) \prime}\
\frac{{\sf P}^{(4) \star} }{{\sf P}^{(6) \star}}\ {\sf A}^{(2) \prime}\ .
\end{eqnarray}
The matrix 
${\sf A}^{(2) \prime}$ is the generalized \{2\}-inverse of the matrix
${\sf P}^{(4) \star} /{\sf P}^{(6) \star}$ (cf., e.g.,
\cite{beni}, Chap.\ 1, p.\ 7).\\

In analogy to the procedure applied in Subsect.\ 2.2, 
we can now successively solve the eqs.\ (\ref{NC6})-(\ref{NC8}).
Choosing $\det{\sf A}^{(2) \prime}\not=0$ (by virtue of eq.\ (\ref{NC10})
this entails
$\det{\sf P}^{(4) \star} \not=0$), we immediately find from eq.\
(\ref{NC10}) an explicit expression for ${\sf A}^{(2) \prime}$.
This can be inserted into eq.\ (\ref{NC7}) to also find
an explicit expression for ${\sf A}^{(4) \star\prime}$. 
Finally, both these explicit expressions for ${\sf A}^{(2) \prime}$
and ${\sf A}^{(4) \star\prime}$ can now be inserted into
eq.\ (\ref{NC8}) to solve it for $A_{1 2 3,1 2 3}^{(6) \prime}$.
The results obtained read as follows.
\begin{eqnarray}
\label{NC11}
{\sf A}^{(2) \prime} &=&
{\sf P}^{(6) \star}\ \left[{\sf P}^{(4) \star}\right]^{-1}\ =\
\frac{{\sf P}^{(6) \star}}{\det {\sf P}^{(4) \star}}\ 
{\rm adj}\; {\sf P}^{(4) \star}\\[0.3cm]
\label{NC12}
{\sf A}^{(4) \star\prime} &=&
- \frac{\left({\sf P}^{(6) \star}\right)^2}{\det {\sf P}^{(4) \star}}
\left[\frac{{\sf P}^{(6) \star}}{\det {\sf P}^{(4) \star}}\
{\sf P}^{(4) \star} {\sf A}^{(2)} {\sf P}^{(4) \star}\ -\ 
{\sf P}^{(4) \star} \right]\\[0.3cm]
\label{NC13}
A_{1 2 3,1 2 3}^{(6) \prime} &=&
\frac{\left({\sf P}^{(6) \star}\right)^5}{
\left(\det {\sf P}^{(4) \star} \right)^2}
\left\{ 1\ -\ 
\frac{2}{\det {\sf P}^{(4) \star}}\
{\rm tr}\left[{\rm adj}\left({\sf P}^{(4) \star} {\sf A}^{(2)}\right)\right] \right\}
\nonumber\\[0.3cm]
&&\ +\
\frac{3 \left({\sf P}^{(6) \star}\right)^4}{
\left(\det {\sf P}^{(4) \star}\right)^2}\ 
{\rm tr}\left( {\sf P}^{(4) \star} {\sf A}^{(2)} \right)
\ -\ 
\frac{4 \left({\sf P}^{(6) \star}\right)^3}{\det {\sf P}^{(4) \star}}
\end{eqnarray}
In deriving eq.\ (\ref{NC13}) we have made use of the identity
(\ref{B2}) given in Appendix B.
In analogy to the eqs.\ (\ref{NC3c}) and (\ref{NC4}), 
we can now define
\begin{eqnarray}
\label{NC14b}
{\sf P}^{(4) \star\prime}&=&{\rm adj}\; {\sf A}^{(2) \prime}\ -\ 
{\sf A}^{(4) \star\prime}\ ,\\[0.3cm]
\label{NC14}
{\sf P}^{(6) \star\prime}&=&
\det{\sf A}^{(2) \prime}\ -\ 
{\rm tr}\left({\sf A}^{(4) \star\prime} {\sf A}^{(2) \prime}\right)
-\ {\sf A}^{(6) \star\prime}\ ,
\end{eqnarray}
and from the eqs.\ (\ref{NC11})-(\ref{NC13}) we find
\begin{eqnarray}
\label{NC15}
{\sf P}^{(4) \star\prime} &=&
\frac{\left({\sf P}^{(6) \star}\right)^3}{
\left(\det {\sf P}^{(4) \star}\right)^2}\
{\sf P}^{(4) \star} {\sf A}^{(2)} {\sf P}^{(4) \star}\ ,\\[0.3cm]
\label{NC16}
{\sf P}^{(6) \star\prime} &=&
- \frac{\left({\sf P}^{(6) \star}\right)^5}{
\left(\det {\sf P}^{(4) \star} \right)^2}
\left\{ 1\ -\ 
\frac{2}{\det {\sf P}^{(4) \star}}\
{\rm tr}\left[{\rm adj}\left({\sf P}^{(4) \star} {\sf A}^{(2)}\right)\right] \right\}
\nonumber\\[0.3cm]
&&\ -\
\frac{2 \left({\sf P}^{(6) \star}\right)^4}{
\left(\det {\sf P}^{(4) \star}\right)^2}\ 
{\rm tr}\left( {\sf P}^{(4) \star} {\sf A}^{(2)} \right)
\ +\ 
\frac{2 \left({\sf P}^{(6) \star}\right)^3}{
\det {\sf P}^{(4) \star}}\ .
\end{eqnarray}
Taking the determinant on both sides of the eqs.\ (\ref{NC11}) and
(\ref{NC15}) provides us
with the following useful relations.
\begin{eqnarray}
\label{NC17}
\det {\sf A}^{(2) \prime} &=&
\frac{\left({\sf P}^{(6) \star}\right)^3}{\det {\sf P}^{(4) \star}}\\[0.3cm]
\label{NC18}
\det {\sf P}^{(4) \star\prime} &=&
\frac{\left({\sf P}^{(6) \star}\right)^9}{
\left(\det {\sf P}^{(4) \star}\right)^4}\
\det {\sf A}^{(2)}
\end{eqnarray}

Finally, also for the case $n = 3$ we derive equations which
describe the inverse of the action map $f$ (The comment
made in footnote \ref{footn3} of Subsect.\ 2.2 also applies here.).
From eqs.\ (\ref{NC11}), (\ref{NC12}), (\ref{NC15}) we find 
\begin{eqnarray}
\label{NC19}{\sf P}^{(4) \star} &=&
{\sf P}^{(6) \star}\ \left[{\sf A}^{(2) \prime}\right]^{-1}\ =\
\frac{{\sf P}^{(6) \star}}{\det {\sf A}^{(2) \prime}}\ 
{\rm adj}\; {\sf A}^{(2) \prime}\ ,\\[0.3cm]
\label{NC20}
{\sf A}^{(4) \star} &=&
\frac{{\sf P}^{(6) \star}}{\det {\sf A}^{(2) \prime}}
\left\{\frac{{\sf P}^{(6) \star}}{\left(\det {\sf A}^{(2) \prime}\right)^3}\
{\rm adj} \left( {\sf A}^{(2) \prime} {\sf P}^{(4) \star\prime}
 {\sf A}^{(2) \prime}\right)\ -\ 
{\rm adj}\; {\sf A}^{(2) \prime} \right\}\ ,\\[0.3cm]
\label{NC21}
{\sf A}^{(2)} &=&
\frac{{\sf P}^{(6) \star}}{\left(\det {\sf A}^{(2) \prime}\right)^2}\
{\sf A}^{(2) \prime} {\sf P}^{(4) \star\prime} {\sf A}^{(2) \prime}
\end{eqnarray}
where now ${\sf P}^{(6) \star}$ is being understood as a
function of the primed quantities whose explicit shape remains
to be determined. Inserting eqs.\ (\ref{NC19}), (\ref{NC20})
into eq.\ (\ref{NC16}) allows us to derive the following 
explicit representation
of the partition function ${\sf P}^{(6) \star}$ in terms of the
primed quantities.
\begin{eqnarray}
\label{NC22}
{\sf P}^{(6) \star}&=&\left(\det {\sf A}^{(2) \prime}\right)^2
\Bigg\{ 2\ \det {\sf A}^{(2) \prime}\ -\ 
2\ {\rm tr}\left( {\sf P}^{(4) \star\prime} {\sf A}^{(2) \prime} \right)
\nonumber\\[0.3cm]
&&\ +\ \frac{2}{\det {\sf A}^{(2) \prime}}\
{\rm tr}\left[{\rm adj}\left({\sf P}^{(4) \star\prime} 
{\sf A}^{(2) \prime}\right)\right]\ -\ 
{\sf P}^{(6) \star\prime}\Bigg\}^{-1}
\end{eqnarray}
In principle, on the basis of this result also an explicit representation 
of $A_{1 2 3,1 2 3}^{(6)}$ in terms of the primed quantities
can be established (relying on eq.\ (\ref{NC4})) but we refrain 
from also writing it down here. As one recognizes from eq.\ (\ref{NC22}),
in the case $n=3$ the description of the inverse of the action map $f$ 
already involves fairly complicated expressions and we will not
attempt to generalize these in the next subsection to the case
$n=4$.\\

The results obtained in the present subsection can be checked 
for consistency in two ways. First, based on the procedure 
described in the Introduction in the context of eqs.\ (\ref{M9a1}),
(\ref{M9a2}) one can convince oneself that the results -- wherever
appropriate -- are consistent with the results obtained in Subsect.\
2.2 for the case of the Grassmann algebra ${\cal G}_4$ ($n=2$). 
Second, choosing for $G_0[\{\bar\Psi\},\{\Psi\} ]$ the form
(\ref{M18}) one can also convince oneself that then 
${\sf A}^{(2) \prime} = {\sf A}^{(2)}$ and
${\sf A}^{(4) \star\prime}$, $A^{(6) \prime}_{1 2 3, 1 2 3}$ 
vanish as expected.\\

\subsection{Explicit calculation: $n=4$}

We are now prepared to study the algebraically most involved
case to be treated in the present article -- the case of the 
Grassmann algebra ${\cal G}_8$ ($n=4$). The calculational
experience collected in the last two subsections allows us to 
manage the fairly involved expressions.\\

We start again by parametrizing $G_0$ and $G$ according to our
general Ansatz (cf.\ eq.\ (\ref{M9})).
\begin{eqnarray}
\label{ND1}
G_0[\{\bar\Psi\},\{\Psi\} ]
&=&A^{(0)}\ +\ \sum_{l,m=1}^4 A_{l,m}^{(2)} 
\bar\Psi_l \Psi_m\nonumber\\[0.3cm]
&&\ +\ \frac{1}{4}
\sum_{l_1,l_2,m_1,m_2=1}^4 A_{l_1 l_2,m_1 m_2}^{(4)} 
\bar\Psi_{l_1}\bar\Psi_{l_2} \Psi_{m_1}\Psi_{m_2}\ \ \ \ \ \ \ \
\nonumber\\[0.3cm]
&&\ +\ \frac{1}{36}
\sum_{l_1,l_2,l_3,m_1,m_2,m_3=1}^4 
A_{l_1 l_2 l_3,m_1 m_2 m_3}^{(6)}
\bar\Psi_{l_1}\bar\Psi_{l_2}\bar\Psi_{l_3} 
\Psi_{m_1}\Psi_{m_2}\Psi_{m_3}\nonumber\\[0.3cm]
&&\ +\ A_{1 2 3 4,1 2 3 4}^{(8)} 
\bar\Psi_1\bar\Psi_2\bar\Psi_3\bar\Psi_4 \Psi_1\Psi_2\Psi_3\Psi_4
\end{eqnarray}
For $G$ the analogous representation can be used.
\begin{eqnarray}
\label{ND2}
G[\{\bar\Psi\},\{\Psi\} ]
&=&A^{(0) \prime}\ +\ \sum_{l,m=1}^4 A_{l,m}^{(2) \prime} 
\bar\Psi_l \Psi_m\nonumber\\[0.3cm]
&&\ +\ \frac{1}{4}
\sum_{l_1,l_2,m_1,m_2=1}^4 A_{l_1 l_2,m_1 m_2}^{(4) \prime} 
\bar\Psi_{l_1}\bar\Psi_{l_2} \Psi_{m_1}\Psi_{m_2}\ \ \ \ \ \ \ \
\nonumber\\[0.3cm]
&&\ +\ \frac{1}{36}
\sum_{l_1,l_2,l_3,m_1,m_2,m_3=1}^4 
A_{l_1 l_2 l_3,m_1 m_2 m_3}^{(6) \prime}
\bar\Psi_{l_1}\bar\Psi_{l_2}\bar\Psi_{l_3} 
\Psi_{m_1}\Psi_{m_2}\Psi_{m_3}\nonumber\\[0.3cm]
&&\ +\ A_{1 2 3 4,1 2 3 4}^{(8) \prime}
\bar\Psi_1\bar\Psi_2\bar\Psi_3\bar\Psi_4 \Psi_1\Psi_2\Psi_3\Psi_4
\end{eqnarray}
The partition function reads (cf.\ eq.\ (\ref{M14b}))
\begin{eqnarray}
\label{ND4}
P&=& {\rm e}^{\displaystyle\ A^{(0) \prime}}\ =\ 
{\sf P}^{(8) \star}\nonumber\\[0.3cm]
&=& \det{\sf A}^{(2)}\ -\
{\rm tr}\left[{\sf A}^{(4) \star} C_2\left({\sf A}^{(2)}\right)\right]
\ +\ \frac{1}{2}\ {\rm tr}\left({\sf A}^{(4) \star} {\sf A}^{(4)}\right)
\nonumber\\[0.3cm]
&&\ -\ {\rm tr}\left({\sf A}^{(6) \star} {\sf A}^{(2)}\right)
+\ {\sf A}^{(8) \star}\\[0.3cm]
\label{ND4b}
&=&6\ \det{\sf A}^{(2)}\ -\ 2\ 
{\rm tr}\left[{\sf P}^{(4) \star} C_2\left({\sf A}^{(2)}\right)\right]
\ +\ \frac{1}{2}\ {\rm tr}\left({\sf  P}^{(4) \star} {\sf P}^{(4)}\right)
\nonumber\\[0.3cm]
&&\ +\ 
{\rm tr}\left({\sf P}^{(6) \star} {\sf A}^{(2)}\right)\ +\ 
{\sf A}^{(8) \star}
\end{eqnarray}
In analogy to eq.\ (\ref{NB3}), 
here ${\sf A}^{(8) \star} = A_{1 2 3 4,1 2 3 4}^{(8)}$ applies.
In the lower line (eq.\ (\ref{ND4b})), we have made use of the expressions
(cf.\ eqs.\ (\ref{NC3c}), (\ref{NC4}))
\begin{eqnarray}
\label{ND3h}
{\sf P}^{(4) \star}&=&
C_2\left({\sf A}^{(2)}\right)^\star\ -\ {\sf A}^{(4) \star}\ ,\\[0.3cm]
\label{ND3d}
{\sf P}^{(6) \star}&=&{\rm adj}\; {\sf A}^{(2)}
\ -\ {\sf F}_a\left({\sf A}^{(2)},{\sf A}^{(4)}\right) \ -\ 
{\sf A}^{(6) \star}
\end{eqnarray}
$\left({\rm adj}\; {\sf A}^{(2)}\ = 
\ C_3\left({\sf A}^{(2)}\right)^\star\right)$. The form ${\sf F}_a$
is defined as follows.
\begin{eqnarray}
\label{ND3b}
{\sf F}_a\left({\sf A}^{(2)},{\sf A}^{(4)}\right)_{lm}
&=&\epsilon_{l r K}\ \epsilon_{m s N}\
{\sf A}^{(2)}_{s r}\ {\sf A}^{(4)}_{N K}
\end{eqnarray}
In making the transition from eq.\
(\ref{ND4}) to eq.\ (\ref{ND4b}) we have used the relations 
\begin{eqnarray}
\label{ND20a}
2\ {\rm tr}\left[ C_2\left({\sf A}^{(2)}\right) {\sf A}^{(4) \star}\right]
&=& {\rm tr}\left[{\sf F}_a\left({\sf A}^{(2)},{\sf A}^{(4)}\right)
 {\sf A}^{(2)}\right]\ ,\\[0.3cm]
\label{ND20b}
C_2\left({\sf A}^{(2)}\right) C_2\left({\sf A}^{(2)}\right)^\star
\ =\ C_2\left({\sf A}^{(2)}\right)^\star C_2\left({\sf A}^{(2)}\right)
&=& \det {\sf A}^{(2)}\ {\bf 1}_6
\end{eqnarray}
(Eq.\ (\ref{ND20b}) is a special case of eq.\ (\ref{A2}), see Appendix A.).
As next step, we can calculate 
$W[\{\bar\eta\},\{\eta\}]$ which reads (To arrive at it we only assume
${\sf P}^{(8) \star}\not= 0$.)
\begin{eqnarray}
\label{ND5}
W[\{\bar\eta\},\{\eta\}] &=& \ln {\sf P}^{(8) \star}\ -\
\frac{{\sf P}_{lm}^{(6) \star}}{{\sf P}^{(8) \star}}\
\bar\eta_l \eta_m\nonumber\\[0.3cm] 
&&-\ \frac{{\sf P}_{LM}^{(4) \star}}{{\sf P}^{(8) \star}}\ 
\bar\eta_{l_1}\bar\eta_{l_2}\eta_{m_1}\eta_{m_2}\
-\ \frac{1}{2}\ 
\left(\frac{{\sf P}_{lm}^{(6) \star}}{
{\sf P}^{(8) \star}}\ \bar\eta_l \eta_m\right)^2\nonumber\\[0.3cm]
&&+\ \frac{1}{{\sf P}^{(8) \star}}\ 
\Bigg[{\sf A}^{(2) \star}\ -\ 
\frac{{\sf F}_a\left({\sf P}^{(6) \star},{\sf P}^{(4) \star}\right)^\star
}{{\sf P}^{(8) \star}}\nonumber\\[0.3cm]
&&\ +\ \frac{2\ C_3\left( {\sf P}^{(6) \star}\right)}{
\left({\sf P}^{(8) \star}\right)^2}
\Bigg]_{LM}\ \bar\eta_{l_1}\bar\eta_{l_2}\bar\eta_{l_3} 
\eta_{m_1}\eta_{m_2}\eta_{m_3}\nonumber\\[0.3cm]
&&+\ \frac{1}{{\sf P}^{(8) \star}}\ 
\Bigg\{1\ -\ 
\frac{{\rm tr}\left({\sf P}^{(6) \star} {\sf A}^{(2)}\right)}{
{\sf P}^{(8) \star}}
\ -\ 
\frac{{\rm tr}\left({\sf P}^{(4) \star} {\sf P}^{(4)}\right)}{2\ 
{\sf P}^{(8) \star}}\nonumber\\[0.3cm]
&&\ +\ \frac{2\ 
{\rm tr}\left[{\sf P}^{(4)} C_2\left({\sf P}^{(6) \star}\right)\right]
}{\left({\sf P}^{(8) \star}\right)^2}
-\ \frac{6\det {\sf P}^{(6) \star}}{
\left({\sf P}^{(8) \star}\right)^3}\Bigg\}\
\bar\eta_1\bar\eta_2\bar\eta_3\bar\eta_4 \eta_1\eta_2\eta_3\eta_4\ \ \ \ .
\ \ \ \ \ \
\end{eqnarray}

In the  following, we need a number of forms which we list here
for further reference. The index convention applied
here requires some explanation. For example,
$\left({\sf A}^{(4) \star\prime} {\cal E}^{(2)}\right)_{l t u r}$
up to the sign denotes elements of the $6\times 6$ matrix 
${\sf A}^{(4) \star\prime} {\cal E}^{(2)}$. If $l<t$, $u<r$,
it denotes the matrix element 
$\left({\sf A}^{(4) \star\prime} {\cal E}^{(2)}\right)_{\{l,t\}\{u,r\}}$.
If $l>t$, $u<r$,
it denotes the matrix element 
$\left(-{\sf A}^{(4) \star\prime} {\cal E}^{(2)}\right)_{\{t,l\}\{u,r\}}$
and if $l<t$, $u>r$,
it denotes the matrix element 
$\left(-{\sf A}^{(4) \star\prime} {\cal E}^{(2)}\right)_{\{l,t\}\{r,u\}}$.
And finally, if
$l>t$, $u>r$,
it denotes the matrix element 
$\left({\sf A}^{(4) \star\prime} {\cal E}^{(2)}\right)_{\{t,l\}\{r,u\}}$.
Of course, 
$\left({\sf A}^{(4) \star\prime} {\cal E}^{(2)}\right)_{\{l,t\}\{u,r\}} = 0$
if $l=t$ or $u = r$. In the following, summation is understood over
repeated indices.
\begin{eqnarray}
\label{ND21a}
{\sf F}_b\left({\sf A}^{(2)\prime} {\sf P}^{(6) \star}\right)_{LM}
&=&\epsilon_{L r k}\ 
\left({\sf A}^{(2)\prime} {\sf P}^{(6) \star}\right)_{s r}\ \epsilon_{s k M}
\\[0.3cm]
\label{ND21b}
{\sf F}_c\left(
{\sf A}^{(4) \star\prime}, {\sf P}^{(6) \star},{\sf A}^{(4) \star\prime} \right)_{lm}
&=&\left({\sf A}^{(4) \star\prime} {\cal E}^{(2)}\right)_{l t u r}\
{\sf P}^{(6) \star}_{s r}\
\left({\cal E}^{(2)} {\sf A}^{(4) \star\prime}\right)_{s t u m}
\end{eqnarray}
\begin{eqnarray}
\label{ND21c}
&&\hspace{-1.5cm}
{\sf F}_{d1}\left(
{\sf A}^{(4) \star\prime}, {\sf A}^{(2) \prime},
{\sf P}^{(4)} C_2\left({\sf A}^{(2) \prime}\right)^\star \right)_{lm}
\nonumber\\[0.3cm]
&=&\left({\sf A}^{(4) \star\prime} {\cal E}^{(2)}\right)_{l r t u}\
{\sf A}^{(2) \prime}_{s r}\
\left[{\cal E}^{(2)}
{\sf P}^{(4)} C_2\left({\sf A}^{(2) \prime}\right)^\star
\right]_{t s u m}\\[0.3cm]
\label{ND21d}
&&\hspace{-1.5cm}
{\sf F}_{d2}\left(
C_2\left({\sf A}^{(2) \prime}\right)^\star {\sf P}^{(4)},
{\sf A}^{(2) \prime}, {\sf A}^{(4) \star\prime} \right)_{lm}
\nonumber\\[0.3cm]
&=&\left[ C_2\left({\sf A}^{(2) \prime}\right)^\star {\sf P}^{(4)}
{\cal E}^{(2)}\right]_{l u t r}\
{\sf A}^{(2) \prime}_{s r}\
\left({\cal E}^{(2)} {\sf A}^{(4) \star\prime} 
 \right)_{t u s m}\\[0.3cm]
\label{ND21e}
&&\hspace{-1.5cm}
{\sf F}_e\left({\sf A}^{(2) \prime}, {\sf A}^{(4) \prime},{\sf A}^{(4) \prime},
{\sf A}^{(2) \prime}\right)_{LM}\nonumber\\[0.3cm]
&=&{\cal E}^{(2)}_{L a b}\ {\sf A}^{(2) \prime}_{r a} 
\left({\cal E}^{(2)} {\sf A}^{(4) \prime} \right)_{r t b u} 
\left({\sf A}^{(4) \prime} {\cal E}^{(2)} \right)_{d t s u}
{\sf A}^{(2) \prime}_{c s}\ {\cal E}^{(2)}_{c d M}\\[0.3cm]
\label{ND21f}
&&\hspace{-1.5cm}
{\sf F}_f\left( {\sf A}^{(4) \star\prime}, {\sf A}^{(2) \prime}
{\sf P}^{(6) \star},{\sf A}^{(4) \prime} \right)_{lm}
\nonumber\\[0.3cm]
&=&\left({\sf A}^{(4) \star\prime} {\cal E}^{(2)}\right)_{l c d a}\
\left({\sf A}^{(2) \prime} {\sf P}^{(6) \star}\right)_{b a}\
\left({\cal E}^{(2)} {\sf A}^{(4) \prime}\right)_{b d c m}\\[0.3cm]
&&\hspace{-1.5cm}
\label{ND21g}
{\sf F}_g\left(C_2\left({\sf P}^{(6) \star}\right) {\sf P}^{(4)}, 
{\sf P}^{(6) \star},{\sf P}^{(6) \star},
{\sf P}^{(4)} C_2\left({\sf P}^{(6) \star}\right)
\right)_{LM}\nonumber\\[0.3cm]
&=&{\cal E}^{(2)}_{L a b}\ 
\left[{\cal E}^{(2)} C_2\left({\sf P}^{(6) \star}\right)
{\sf P}^{(4)}  \right]_{a r b t}
{\sf P}^{(6) \star}_{r s}\
{\sf P}^{(6) \star}_{t u}
\left[{\sf P}^{(4)} C_2\left({\sf P}^{(6) \star}\right)
{\cal E}^{(2)} \right]_{c s d  u}
\ {\cal E}^{(2)}_{c d M}
\end{eqnarray}

To arrive at the further results it is useful to take note
of the equation
\begin{eqnarray}
\label{ND22}
\left({\sf P}_{lm}^{(6) \star} \bar\eta_l \eta_m\right)^2 &=&
-2\ C_2\left({\sf P}^{(6) \star}\right)_{LM}\ 
\bar\eta_{l_1}\bar\eta_{l_2}\eta_{m_1}\eta_{m_2}
\end{eqnarray}
We now apply exactly the same procedure as in Subsects.\ 2.2, 2.3.
We insert eq.\ (\ref{ND2}) onto the l.h.s.\ of eq.\ (\ref{M7c0})
and the explicit expressions for $\bar\eta$, $\eta$ found from eq.\
(\ref{ND2}) according to eq.\ (\ref{M7b}) on its r.h.s.. Again, comparing
coefficients on both sides we find the following four coupled 
nonlinear matrix equations.
\begin{eqnarray}
\label{ND6}
{\sf A}^{(2) \prime} &=& 2 {\sf A}^{(2) \prime}\ -\ 
{\sf A}^{(2) \prime}\
\frac{{\sf P}^{(6) \star} }{{\sf P}^{(8) \star}}\ {\sf A}^{(2) \prime}
\\[0.3cm]
\label{ND7}
{\sf A}^{(4) \star\prime} &=&
4 {\sf A}^{(4) \star\prime}\ -\ 
\frac{{\sf F}_b\left({\sf A}^{(2)\prime} {\sf P}^{(6) \star}\right)}{
{\sf P}^{(8) \star}}\ {\sf A}^{(4) \star\prime}\ -\ {\sf A}^{(4) \star\prime}\ 
\frac{{\sf F}_b\left({\sf P}^{(6) \star} {\sf A}^{(2)\prime} \right)}{
{\sf P}^{(8) \star}}\nonumber\\[0.3cm]
&&-\ 
C_2\left({\sf A}^{(2)\prime}\right)^\star\ 
\frac{{\sf P}^{(8) \star}\ {\sf P}^{(4)} - 
C_2\left({\sf P}^{(6) \star}\right)^\star}{
\left({\sf P}^{(8) \star}\right)^2}\ 
C_2\left({\sf A}^{(2)\prime}\right)^\star\\[0.3cm]
\label{ND8}
{\sf A}^{(6) \star\prime}&=&
6 {\sf A}^{(6) \star\prime}\ +\ 
\frac{1}{{\sf P}^{(8) \star}}\Bigg\{ {\sf A}^{(6) \star\prime}
\left[{\sf A}^{(2)\prime} {\sf P}^{(6) \star} - 
{\rm tr}\left({\sf A}^{(2)\prime} {\sf P}^{(6) \star}\right) {\bf 1}_4\right]
\nonumber\\[0.3cm]
&&\ + \ \left[{\sf P}^{(6) \star} {\sf A}^{(2)\prime} -
{\rm tr}\left({\sf P}^{(6) \star} {\sf A}^{(2)\prime}\right) {\bf 1}_4\right]
{\sf A}^{(6) \star\prime}
\nonumber\\[0.3cm]
&&\ -\ {\sf F}_c\left(
{\sf A}^{(4) \star\prime}, {\sf P}^{(6) \star},{\sf A}^{(4) \star\prime} \right)
\ +\ {\sf F}_{d1}\left(
{\sf A}^{(4) \star\prime}, {\sf A}^{(2) \prime},
{\sf P}^{(4)} C_2\left({\sf A}^{(2) \prime}\right)^\star \right)
\nonumber\\[0.3cm]
&&\ +\ {\sf F}_{d2}\left(
C_2\left({\sf A}^{(2) \prime}\right)^\star {\sf P}^{(4)},
{\sf A}^{(2) \prime}, {\sf A}^{(4) \star\prime} \right)\ +\ 
{\rm adj}\left( {\sf A}^{(2) \prime}\right) 
{\sf A}^{(2)} {\rm adj}\left( {\sf A}^{(2) \prime}\right)
\Bigg\}\nonumber\\[0.3cm]
&&\ -\ \frac{1}{\left({\sf P}^{(8) \star}\right)^2}\Bigg\{
{\sf F}_{d1}\left(
{\sf A}^{(4) \star\prime}, {\sf A}^{(2) \prime},
C_2\left({\sf A}^{(2) \prime} {\sf P}^{(6) \star} \right)^\star \right)
\nonumber\\[0.3cm]
&&\ +\ {\sf F}_{d2}\left(
C_2\left({\sf P}^{(6) \star} {\sf A}^{(2) \prime}\right)^\star,
{\sf A}^{(2) \prime}, {\sf A}^{(4) \star\prime} \right)
\nonumber\\[0.3cm]
&&\ +\ {\rm adj}\left( {\sf A}^{(2) \prime}\right) 
{\sf F}_a\left({\sf P}^{(6) \star},{\sf P}^{(4) \star}\right)
{\rm adj}\left( {\sf A}^{(2) \prime}\right) \Bigg\}
\nonumber\\[0.3cm]
&&\ +\ \frac{2}{\left({\sf P}^{(8) \star}\right)^3}\ 
{\rm adj}\left( {\sf A}^{(2) \prime} {\sf P}^{(6) \star} {\sf A}^{(2) \prime}
\right)\ \ \ \ \ \\[0.3cm]
\label{ND9}
A_{1 2 3 4,1 2 3 4}^{(8) \prime} &=&
8 A_{1 2 3 4,1 2 3 4}^{(8) \prime}\ +\ 
\frac{1}{{\sf P}^{(8) \star}}\Bigg\{ - 2\ 
A_{1 2 3 4,1 2 3 4}^{(8) \prime}\
{\rm tr}\left( {\sf P}^{(6) \star} {\sf A}^{(2) \prime} \right)
\nonumber\\[0.3cm]
&&\ -\ 2\ {\rm tr}\left[ {\sf P}^{(6) \star}
{\sf F}_a\left({\sf A}^{(6) \star\prime},{\sf A}^{(4) \star\prime}\right)\right]
\nonumber\\[0.3cm]
&&\ -\ 2\ {\rm tr}\left[ {\sf A}^{(2) \prime} {\sf A}^{(6) \star\prime}
{\sf A}^{(2) \prime}
{\sf F}_a\left({\sf A}^{(2) \prime},{\sf P}^{(4)}\right)\right]
\nonumber\\[0.3cm]
&&\ -\ {\rm tr}\left[ {\sf P}^{(4)}\
{\sf F}_e\left({\sf A}^{(2) \prime}, {\sf A}^{(4) \prime},{\sf A}^{(4) \prime},
{\sf A}^{(2) \prime}\right)\right]\nonumber\\[0.3cm]
&&\ -\  \frac{1}{2}\ {\rm tr}\left[
{\sf F}_c\left( {\sf P}^{(4) \star} C_2\left({\sf A}^{(2) \prime}\right),
{\bf 1}_4,{\sf A}^{(4) \star\prime} {\sf A}^{(4) \prime}\right)\right]
\nonumber\\[0.3cm]
&&\ -\  \frac{1}{2}\ {\rm tr}\left[
{\sf F}_c\left({\sf A}^{(4) \prime} {\sf A}^{(4) \star\prime},
{\bf 1}_4,C_2\left({\sf A}^{(2) \prime}\right) {\sf P}^{(4) \star}\right)\right]
\nonumber\\[0.3cm]
&&\ +\ {\rm tr}\left[ {\sf A}^{(4) \star\prime}\
C_2\left({\sf A}^{(2) \prime}\right)\
{\sf F}_b\left(\left({\rm adj}\; {\sf A}^{(2) \prime}\right)
{\sf A}^{(2)}\right)\right]\nonumber\\[0.3cm]
&&\ +\ {\rm tr}\left[ C_2\left({\sf A}^{(2) \prime}\right)\ 
{\sf A}^{(4) \star\prime}\
{\sf F}_b\left({\sf A}^{(2)} {\rm adj}\; {\sf A}^{(2) \prime}\right)
\right]\ +\ \left(\det {\sf A^{(2) \prime}}\right)^2
\Bigg\}\nonumber\\[0.3cm]
&&\ +\ \frac{1}{\left({\sf P}^{(8) \star}\right)^2}\Bigg\{
2\ {\rm tr}\left( {\sf A}^{(2) \prime} {\sf A}^{(6) \star\prime}
{\sf A}^{(2) \prime} {\sf P}^{(6) \star} \right)\
{\rm tr}\left({\sf A}^{(2) \prime} {\sf P}^{(6) \star}
\right)\nonumber\\[0.3cm]
&&\ -\
2\ {\rm tr}\left( {\sf A}^{(2) \prime} {\sf A}^{(6) \star\prime}
{\sf A}^{(2) \prime} {\sf P}^{(6) \star} {\sf A}^{(2) \prime} {\sf P}^{(6) \star}
\right)\nonumber\\[0.3cm]
&&\ +\ {\rm tr}\left[{\sf A}^{(2) \prime} {\sf P}^{(6) \star}
{\sf A}^{(2) \prime}\ {\sf F}_c\left(
{\sf A}^{(4) \star\prime}, {\sf P}^{(6) \star},{\sf A}^{(4) \star\prime} \right)
\right]\nonumber\\[0.3cm]
&&\ -\ {\rm tr}\left[{\sf P}^{(6) \star} {\sf A}^{(2) \prime}\ 
{\sf F}_f\left( {\sf A}^{(4) \star\prime}, {\sf A}^{(2) \prime}
{\sf P}^{(6) \star},{\sf A}^{(4) \prime} \right) \right]\nonumber\\[0.3cm]
&&\ +\ \frac{1}{2}\
{\rm tr}\left[
{\sf F}_c\left( C_2\left({\sf P}^{(6) \star} {\sf A}^{(2) \prime}\right),
{\bf 1}_4,{\sf A}^{(4) \star\prime} {\sf A}^{(4) \prime}\right)\right]
\nonumber\\[0.3cm]
&&\ +\ \frac{1}{2}\
{\rm tr}\left[
{\sf F}_c\left({\sf A}^{(4) \prime} {\sf A}^{(4) \star\prime},
{\bf 1}_4,C_2\left({\sf A}^{(2) \prime} {\sf P}^{(6) \star}\right)\right)\right]
\nonumber\\[0.3cm]
&&\ -\ {\rm tr}\left[
{\sf F}_a\left({\sf P}^{(6) \star},{\sf P}^{(4) \star}\right)\
{\sf F}_a\left({\bf 1}_4,{\sf A}^{(4) \star\prime} 
C_2\left({\sf A}^{(2) \prime}\right)\right)\
{\rm adj}\; {\sf A}^{(2) \prime}\right]\nonumber\\[0.3cm]
&&\ -\ {\rm tr}\left[
{\sf F}_a\left({\bf 1}_4,C_2\left({\sf A}^{(2) \prime}\right)
{\sf A}^{(4) \star\prime} \right)\
{\sf F}_a\left({\sf P}^{(6) \star},{\sf P}^{(4) \star}\right)\
{\rm adj}\; {\sf A}^{(2) \prime}\right]\nonumber\\[0.3cm]
&&\ -\ \left(\det {\sf A^{(2) \prime}}\right)^2\ \left[
{\rm tr}\left({\sf P}^{(6) \star} {\sf A}^{(2)}\right)
\ +\ \frac{1}{2}\
{\rm tr}\left({\sf P}^{(4) \star} {\sf P}^{(4)}\right)\right]
\Bigg\}
\nonumber\\[0.3cm]
&&\ +\ \frac{2}{\left({\sf P}^{(8) \star}\right)^3}\Bigg\{
{\rm tr}\left[ {\sf A}^{(4) \star\prime}\
C_2\left({\sf A}^{(2) \prime}\right)\
{\sf F}_b\left({\rm adj}\; 
\left({\sf P}^{(6) \star} {\sf A}^{(2) \prime}\right)\right)
\right]\nonumber\\[0.3cm]
&&\ +\ {\rm tr}\left[ C_2\left({\sf A}^{(2) \prime}\right)\ 
{\sf A}^{(4) \star\prime}\
{\sf F}_b\left({\rm adj}\; 
\left({\sf A}^{(2) \prime} {\sf P}^{(6) \star} \right)\right)
\right]\nonumber\\[0.3cm]
&&\ +\ \left(\det {\sf A^{(2) \prime}}\right)^2\
{\rm tr}\left[{\sf P}^{(4)} C_2\left({\sf P}^{(6) \star}\right)\right]
\Bigg\}
\nonumber\\[0.3cm]
&&\ -\ \frac{6}{\left({\sf P}^{(8) \star}\right)^4}
\ \left(\det {\sf A^{(2) \prime}}\right)^2\ \det {\sf P^{(6) \star}} 
\end{eqnarray}
Eq.\ (\ref{ND6}) is equivalent to the equation
\begin{eqnarray}
\label{ND10}
{\sf A}^{(2) \prime} &=&{\sf A}^{(2) \prime}\
\frac{{\sf P}^{(6) \star} }{{\sf P}^{(8) \star}}\ {\sf A}^{(2) \prime}
\end{eqnarray}
The matrix 
${\sf A}^{(2) \prime}$ is the generalized \{2\}-inverse of the matrix
${\sf P}^{(6) \star} /{\sf P}^{(8) \star}$ (cf., e.g.,
\cite{beni}, Chap.\ 1, p.\ 7).\\

For solving the eqs.\ (\ref{ND6})-(\ref{ND9}) we apply again
the same method as in Subsects.\ 2.2 and 2.3.
Choosing $\det{\sf A}^{(2) \prime}\not=0$ (By virtue of eq.\ (\ref{ND10})
this entails
$\det{\sf P}^{(6) \star} \not=0$.), we immediately find from eq.\
(\ref{ND10}) an explicit expression for ${\sf A}^{(2) \prime}$.
This can be inserted into eq.\ (\ref{ND7}) to also find
an explicit expression for ${\sf A}^{(4) \star\prime}$. 
\begin{eqnarray}
\label{ND11}
{\sf A}^{(2) \prime} &=&
{\sf P}^{(8) \star}\ \left[{\sf P}^{(6) \star}\right]^{-1}\ =\
\frac{{\sf P}^{(8) \star}}{\det {\sf P}^{(6) \star}}\ 
{\rm adj}\; {\sf P}^{(6) \star}\\[0.3cm]
\label{ND12}
{\sf A}^{(4) \star\prime} &=&
- \frac{\left({\sf P}^{(8) \star}\right)^2}{\det {\sf P}^{(6) \star}}
\left[\frac{{\sf P}^{(8) \star}}{\det {\sf P}^{(6) \star}}\
C_2\left({\sf P}^{(6) \star}\right)
{\sf P}^{(4)} C_2\left({\sf P}^{(6) \star}\right)\ -\ 
C_2\left({\sf P}^{(6) \star}\right) \right]
\end{eqnarray}
To arrive at eq.\ (\ref{ND12}) we have relied on the following
calculation (cf.\ Appendix A, eqs.\ (\ref{A2}), (\ref{A1c})).
\begin{eqnarray}
\label{ND13}
C_2\left({\sf A}^{(2)\prime}\right)^\star
&=&\left({\sf P}^{(8) \star}\right)^2\ 
C_2\left(\left[{\sf P}^{(6) \star}\right]^{-1}\right)^\star\nonumber\\[0.3cm]
&=&\frac{\left({\sf P}^{(8) \star}\right)^2}{\det {\sf P}^{(6) \star}}\ 
C_2\left(\left[{\sf P}^{(6) \star}\right]^{-1}\right)^{-1}\ =\
\frac{\left({\sf P}^{(8) \star}\right)^2}{\det {\sf P}^{(6) \star}}\ 
C_2\left({\sf P}^{(6) \star}\right)
\end{eqnarray}
Having obtained explicit expressions for ${\sf A}^{(2) \prime}$
and ${\sf A}^{(4) \star\prime}$ we can now insert them into
eq.\ (\ref{ND8}) to solve it. We find
\begin{eqnarray}
\label{ND14}
&&\hspace{-1.5cm}{\sf A}^{(6) \star\prime}\nonumber\\[0.3cm] 
&=&\frac{\left({\sf P}^{(8) \star}\right)^5}{
\left(\det {\sf P}^{(6) \star} \right)^2}\ 
{\sf P}^{(6) \star} \Bigg\{  {\sf A}^{(2)} 
\ -\ \frac{1}{2 \det {\sf P}^{(6) \star}}\
{\sf F}_{d1}\left({\sf P}^{(4)}, {\sf P}^{(6) \star},
C_2\left( {\sf P}^{(6) \star}\right) {\sf P}^{(4)} \right)
\nonumber\\[0.3cm]
&&\ -\ \frac{1}{2 \det {\sf P}^{(6) \star}}\ 
{\sf F}_{d2}\left({\sf P}^{(4)} C_2\left( {\sf P}^{(6) \star}\right),
{\sf P}^{(6) \star}, {\sf P}^{(4)} \right) 
\Bigg\} {\sf P}^{(6) \star}\nonumber\\[0.3cm]
&&\ +\
\frac{3 \left({\sf P}^{(8) \star}\right)^4}{
\left(\det {\sf P}^{(6) \star}\right)^2}\ 
{\sf P}^{(6) \star}{\sf F}_a\left({\sf P}^{(6) \star},{\sf P}^{(4) \star}\right)
{\sf P}^{(6) \star}
\ -\ 
\frac{4 \left({\sf P}^{(8) \star}\right)^3}{\det {\sf P}^{(6) \star}}\ 
{\sf P}^{(6) \star}\\[0.3cm]
\label{ND15}
&=&\frac{\left({\sf P}^{(8) \star}\right)^5}{
\left(\det {\sf P}^{(6) \star} \right)^2}\ 
{\sf P}^{(6) \star}\ {\sf A}^{(2)}\ {\sf P}^{(6) \star}\nonumber\\[0.3cm] 
&&\ +\ \frac{\left({\sf P}^{(8) \star}\right)^5}{
\left(\det {\sf P}^{(6) \star} \right)^4}\
{\sf F}_{c}\left(C_2\left( {\sf P}^{(6) \star}\right)
{\sf P}^{(4)} C_2\left( {\sf P}^{(6) \star}\right),{\sf P}^{(6) \star},
C_2\left( {\sf P}^{(6) \star}\right)
{\sf P}^{(4)} C_2\left( {\sf P}^{(6) \star}\right)\right)\nonumber\\[0.3cm]
&&\ +\
\frac{3 \left({\sf P}^{(8) \star}\right)^4}{
\left(\det {\sf P}^{(6) \star}\right)^2}\ 
{\sf P}^{(6) \star}{\sf F}_a\left({\sf P}^{(6) \star},{\sf P}^{(4) \star}\right)
{\sf P}^{(6) \star}
\ -\ 
\frac{4 \left({\sf P}^{(8) \star}\right)^3}{\det {\sf P}^{(6) \star}}\ 
{\sf P}^{(6) \star}\ .
\end{eqnarray}
The equivalence of eqs.\ (\ref{ND14}) and (\ref{ND15}) is based on the
relation
\begin{eqnarray}
\label{ND16}
&&\hspace{-1.5cm}\left(\det{\sf P}^{(6) \star}\right)\
{\sf P}^{(6) \star}\ {\sf F}_{d1}\left({\sf P}^{(4)}, {\sf P}^{(6) \star},
C_2\left( {\sf P}^{(6) \star}\right) {\sf P}^{(4)} \right)\
{\sf P}^{(6) \star}\nonumber\\[0.3cm]
&=&\left(\det{\sf P}^{(6) \star}\right)\
{\sf P}^{(6) \star}\ 
{\sf F}_{d2}\left({\sf P}^{(4)} C_2\left( {\sf P}^{(6) \star}\right),
{\sf P}^{(6) \star}, {\sf P}^{(4)} \right)\ {\sf P}^{(6) \star}\nonumber\\[0.3cm]
&=& -\
{\sf F}_{c}\left(C_2\left( {\sf P}^{(6) \star}\right)
{\sf P}^{(4)} C_2\left( {\sf P}^{(6) \star}\right),{\sf P}^{(6) \star},
C_2\left( {\sf P}^{(6) \star}\right)
{\sf P}^{(4)} C_2\left( {\sf P}^{(6) \star}\right)\right)\ .
\end{eqnarray}
Finally, inserting eqs.\ (\ref{ND11}), (\ref{ND12}), (\ref{ND15})
into (\ref{ND9}) allows us to find the following explicit
solution for $A_{1 2 3 4,1 2 3 4}^{(8) \prime}$.
\begin{eqnarray}
\label{ND17}
&&\hspace{-1.5cm}A_{1 2 3 4,1 2 3 4}^{(8) \prime}\nonumber\\[0.3cm]
&=&\frac{\left({\sf P}^{(8) \star}\right)^7}{
\left(\det {\sf P}^{(6) \star}\right)^2}\
\left\{ 1\ -\ 2\ {\rm tr}\left[{\sf A}^{(2)}\ 
{\sf F}_a\left(({\sf P}^{(6) \star})^{-1},{\sf P}^{(4)}\right)\right]\right\}
\nonumber\\[0.3cm]
&&\ +\ \frac{\left({\sf P}^{(8) \star}\right)^7}{
\left(\det {\sf P}^{(6) \star}\right)^4}\ 
\Bigg\{ 
{\rm tr}\left[{\sf P}^{(4)}\
{\sf F}_g\left(C_2\left({\sf P}^{(6) \star}\right) {\sf P}^{(4)}, 
{\sf P}^{(6) \star},{\sf P}^{(6) \star},
{\sf P}^{(4)} C_2\left({\sf P}^{(6) \star}\right)\right)\right]
\nonumber\\[0.3cm]
&&\ -\ \frac{1}{2}\
{\rm tr}\left[{\sf F}_c\left( 
C_2\left( {\sf P}^{(6) \star}\right) {\sf P}^{(4)} {\sf P}^{(4) \star}
C_2\left( {\sf P}^{(6) \star}\right)^\star, 
{\bf 1}_4, C_2\left( {\sf P}^{(6) \star} \right) {\sf P}^{(4)}
\right)\right]
\nonumber\\[0.3cm]
&&\ -\ \frac{1}{2}\
{\rm tr}\left[
{\sf F}_c\left( C_2\left( {\sf P}^{(6) \star}\right)^\star {\sf P}^{(4) \star} 
{\sf P}^{(4)} C_2\left( {\sf P}^{(6) \star}\right),{\bf 1}_4,
{\sf P}^{(4)} C_2\left( {\sf P}^{(6) \star}\right)\right)\right]
\Bigg\}\nonumber\\[0.3cm]
&&\ +\ \frac{\left({\sf P}^{(8) \star}\right)^6}{
\left(\det {\sf P}^{(6) \star}\right)^2}\
\Bigg\{
\frac{11}{2}\ {\rm tr}\left({\sf P}^{(4)} {\sf P}^{(4) \star}\right)\ +\ 
5\ {\rm tr}\left({\sf P}^{(6) \star} {\sf A}^{(2)}\right)
\nonumber\\[0.3cm]
&&\ -\ 
\frac{5}{\det {\sf P}^{(6) \star}}\ 
{\rm tr}\left[
{\sf F}_a\left({\bf 1}_4,C_2\left({\sf P}^{(6) \star}\right) {\sf P}^{(4)}\right)
{\sf F}_a\left({\bf 1}_4,{\sf P}^{(4) \star} 
C_2\left({\sf P}^{(6) \star}\right)^\star\right)
\right]\Bigg\}
\nonumber\\[0.3cm]
&&\ +\ 
18\ \frac{\left({\sf P}^{(8) \star}\right)^5}{
\left(\det {\sf P}^{(6) \star}\right)^2}\
{\rm tr}\left[{\sf P}^{(4)} C_2\left({\sf P}^{(6) \star}\right)\right]\ -\ 
30\ \frac{\left({\sf P}^{(8) \star}\right)^4}{
\det {\sf P}^{(6) \star}}
\end{eqnarray}

In analogy to the eqs.\ (\ref{ND3h}), (\ref{ND3d}) and (\ref{ND4}), 
we can now define
\begin{eqnarray}
\label{ND18}
{\sf P}^{(4) \star\prime}&=&
C_2\left({\sf A}^{(2) \prime}\right)^\star\ -\ 
{\sf A}^{(4) \star\prime}\ ,\\[0.3cm]
\label{ND19}
{\sf P}^{(6) \star\prime}&=&{\rm adj}\; {\sf A}^{(2) \prime}
\ -\ {\sf F}_a\left({\sf A}^{(2) \prime},{\sf A}^{(4) \prime}\right) \ -\ 
{\sf A}^{(6) \star\prime}\ ,\\[0.3cm]
\label{ND20}
{\sf P}^{(8) \star\prime}
&=& \det{\sf A}^{(2) \prime}\ -\
{\rm tr}\left[{\sf A}^{(4) \star\prime} 
C_2\left({\sf A}^{(2) \prime}\right)\right]
\ +\ \frac{1}{2}\ {\rm tr}\left({\sf A}^{(4) \star\prime} 
{\sf A}^{(4) \prime}\right)
\nonumber\\[0.3cm]
&&\ -\ {\rm tr}\left({\sf A}^{(6) \star\prime} 
{\sf A}^{(2) \prime}\right)
+\ {\sf A}^{(8) \star\prime}\ ,
\end{eqnarray}
and from eqs.\ (\ref{ND11}), (\ref{ND12}), 
(\ref{ND15}), (\ref{ND17}), we find
\begin{eqnarray}
\label{ND21}
{\sf P}^{(4) \star\prime} &=&
\frac{\left({\sf P}^{(8) \star}\right)^3}{
\left(\det {\sf P}^{(6) \star}\right)^2}\
C_2\left({\sf P}^{(6) \star}\right)
{\sf P}^{(4)} C_2\left({\sf P}^{(6) \star}\right)\ ,\\[0.3cm]
\label{ND22a}
{\sf P}^{(6) \star\prime}
&=&-\ \frac{\left({\sf P}^{(8) \star}\right)^5}{
\left(\det {\sf P}^{(6) \star} \right)^2}\ 
{\sf P}^{(6) \star} \Bigg\{  {\sf A}^{(2)} 
\ -\ \frac{1}{2 \det {\sf P}^{(6) \star}}\
{\sf F}_{d1}\left({\sf P}^{(4)}, {\sf P}^{(6) \star},
C_2\left( {\sf P}^{(6) \star}\right) {\sf P}^{(4)} \right)
\nonumber\\[0.3cm]
&&\ -\ \frac{1}{2 \det {\sf P}^{(6) \star}}\ 
{\sf F}_{d2}\left({\sf P}^{(4)} C_2\left( {\sf P}^{(6) \star}\right),
{\sf P}^{(6) \star}, {\sf P}^{(4)} \right) 
\Bigg\} {\sf P}^{(6) \star}\nonumber\\[0.3cm]
&&\ -\
\frac{2 \left({\sf P}^{(8) \star}\right)^4}{
\left(\det {\sf P}^{(6) \star}\right)^2}\ 
{\sf P}^{(6) \star}{\sf F}_a\left({\sf P}^{(6) \star},{\sf P}^{(4) \star}\right)
{\sf P}^{(6) \star}
\ +\ 
\frac{2 \left({\sf P}^{(8) \star}\right)^3}{
\det {\sf P}^{(6) \star}}\ 
{\sf P}^{(6) \star}\\[0.3cm]
\label{ND22b}
&=&-\ \frac{\left({\sf P}^{(8) \star}\right)^5}{
\left(\det {\sf P}^{(6) \star} \right)^2}\ 
{\sf P}^{(6) \star}\ {\sf A}^{(2)}\ {\sf P}^{(6) \star}\nonumber\\[0.3cm] 
&&\ -\ \frac{\left({\sf P}^{(8) \star}\right)^5}{
\left(\det {\sf P}^{(6) \star} \right)^4}\ \nonumber\\[0.3cm] 
&&\ \ \ \ \ \ \ \ \ 
{\sf F}_{c}\left(C_2\left( {\sf P}^{(6) \star}\right)
{\sf P}^{(4)} C_2\left( {\sf P}^{(6) \star}\right),{\sf P}^{(6) \star},
C_2\left( {\sf P}^{(6) \star}\right)
{\sf P}^{(4)} C_2\left( {\sf P}^{(6) \star}\right)\right)\nonumber\\[0.3cm]
&&\ -\
\frac{2 \left({\sf P}^{(8) \star}\right)^4}{
\left(\det {\sf P}^{(6) \star}\right)^2}\ 
{\sf P}^{(6) \star}{\sf F}_a\left({\sf P}^{(6) \star},{\sf P}^{(4) \star}\right)
{\sf P}^{(6) \star}
\ +\ 
\frac{2 \left({\sf P}^{(8) \star}\right)^3}{\det {\sf P}^{(6) \star}}\ 
{\sf P}^{(6) \star}\ ,\\[0.3cm]
\label{ND23}
{\sf P}^{(8) \star\prime}
&=&\frac{\left({\sf P}^{(8) \star}\right)^7}{
\left(\det {\sf P}^{(6) \star}\right)^2}\
\left\{ 1\ -\ 2\ {\rm tr}\left[{\sf A}^{(2)}\ 
{\sf F}_a\left(({\sf P}^{(6) \star})^{-1},{\sf P}^{(4)}\right)\right]\right\}
\nonumber\\[0.3cm]
&&\ +\ \frac{\left({\sf P}^{(8) \star}\right)^7}{
\left(\det {\sf P}^{(6) \star}\right)^4}\ 
\Bigg\{ 
{\rm tr}\left[{\sf P}^{(4)}\
{\sf F}_g\left(C_2\left({\sf P}^{(6) \star}\right) {\sf P}^{(4)}, 
{\sf P}^{(6) \star},{\sf P}^{(6) \star},
{\sf P}^{(4)} C_2\left({\sf P}^{(6) \star}\right)\right)\right]
\nonumber\\[0.3cm]
&&\ -\ \frac{1}{2}\
{\rm tr}\left[{\sf F}_c\left( 
C_2\left( {\sf P}^{(6) \star}\right) {\sf P}^{(4)} {\sf P}^{(4) \star}
C_2\left( {\sf P}^{(6) \star}\right)^\star, 
{\bf 1}_4, C_2\left( {\sf P}^{(6) \star} \right) {\sf P}^{(4)}
\right)\right]
\nonumber\\[0.3cm]
&&\ -\ \frac{1}{2}\
{\rm tr}\left[
{\sf F}_c\left( C_2\left( {\sf P}^{(6) \star}\right)^\star {\sf P}^{(4) \star} 
{\sf P}^{(4)} C_2\left( {\sf P}^{(6) \star}\right),{\bf 1}_4,
{\sf P}^{(4)} C_2\left( {\sf P}^{(6) \star}\right)\right)\right]
\Bigg\}\nonumber\\[0.3cm]
&&\ +\ 4\ \frac{\left({\sf P}^{(8) \star}\right)^6}{
\left(\det {\sf P}^{(6) \star}\right)^2}\
\Bigg\{
{\rm tr}\left({\sf P}^{(4)} {\sf P}^{(4) \star}\right)\ +\ 
{\rm tr}\left({\sf P}^{(6) \star} {\sf A}^{(2)}\right)
\nonumber\\[0.3cm]
&&\ -\ 
\frac{1}{\det {\sf P}^{(6) \star}}\ 
{\rm tr}\left[
{\sf F}_a\left({\bf 1}_4,C_2\left({\sf P}^{(6) \star}\right) {\sf P}^{(4)}\right)
{\sf F}_a\left({\bf 1}_4,{\sf P}^{(4) \star} 
C_2\left({\sf P}^{(6) \star}\right)^\star\right)
\right]\Bigg\}
\nonumber\\[0.3cm]
&&\ +\ 
12\ \frac{\left({\sf P}^{(8) \star}\right)^5}{
\left(\det {\sf P}^{(6) \star}\right)^2}\
{\rm tr}\left[{\sf P}^{(4)} C_2\left({\sf P}^{(6) \star}\right)\right]\ -\ 
16\ \frac{\left({\sf P}^{(8) \star}\right)^4}{\det {\sf P}^{(6) \star}}\ .
\end{eqnarray}
Taking the determinant on both sides of the eqs.\ (\ref{ND11}) and
(\ref{ND21}) provides us
with the following useful relations.
\begin{eqnarray}
\label{ND24}
\det {\sf A}^{(2) \prime} &=&
\frac{\left({\sf P}^{(8) \star}\right)^4}{\det {\sf P}^{(6) \star}}
\\[0.3cm]
\label{ND25}
\det {\sf P}^{(4) \star\prime}&=&
\frac{\left({\sf P}^{(8) \star}\right)^{18}}{
\left(\det {\sf P}^{(6) \star}\right)^6}\
\det {\sf P}^{(4) \star}
\end{eqnarray}
In deriving eq.\ (\ref{ND25}) we have relied on the following
(Sylvester-Franke) identity (cf.\ Appendix A, eq.\ (\ref{A4})).
\begin{eqnarray}
\det C_2\left({\sf P}^{(6) \star}\right) &=& 
\left(\det {\sf P}^{(6) \star}\right)^3
\end{eqnarray}

We can finally check the obtained results 
for consistency in the same way as done at the end
of the previous subsection for $n=3$. First, based on the procedure 
described in the Introduction in the context of eqs.\ (\ref{M9a1}),
(\ref{M9a2}) one can convince oneself again that the results -- wherever
appropriate -- are consistent with the results obtained in Subsect.\
2.3 for the case of the Grassmann algebra ${\cal G}_6$ ($n=3$). 
Second, choosing for $G_0[\{\bar\Psi\},\{\Psi\} ]$ the form
(\ref{M18}) one can also convince oneself that then 
${\sf A}^{(2) \prime} = {\sf A}^{(2)}$ and
${\sf A}^{(4) \star\prime}$, ${\sf A}^{(6) \star\prime}$,
$A^{(8) \prime}_{1 2 3 4, 1 2 3 4}$ 
vanish as expected. Given the combinatorial factors involved,
this represents a fairly sensitive check of the present results.\\

\subsection{Some heuristics for arbitrary $n$}

Having gained a fairly broad calculational and structural experience
in the previous subsections
in considering the present formalism for the case of
the Grassmann algebras ${\cal G}_{2n}$, $n = 2, 3, 4$,
we are going to generalize now some of the obtained results to
arbitrary values of $n$. This analytic extrapolation is a 
heuristic procedure with heuristic purposes. No proof is 
being attempted here which would need to be the subject of a separate
study.\\

From eqs.\ (\ref{NB8}), (\ref{NC11}), (\ref{ND11}) and
(\ref{NB9}), (\ref{NC12}), (\ref{ND12}) we infer the following
general (for arbitrary values of $n$) form of the matrices
${\sf A}^{(2) \prime}$, ${\sf A}^{(4) \prime}$
(Of course, the result for ${\sf A}^{(2) \prime}$ is elementary.).
\begin{eqnarray}
\label{NE1}
{\sf A}^{(2) \prime} &=&
{\sf P}^{(2n) \star}\ \left[{\sf P}^{(2n - 2) \star}\right]^{-1}\ =\
\frac{{\sf P}^{(2n) \star}}{\det {\sf P}^{(2n -2) \star}}\ 
{\rm adj}\; {\sf P}^{(2n -2 ) \star}\\[0.3cm]
\label{NE2}
{\sf A}^{(4) \prime} &=&
- \frac{\left({\sf P}^{(2n) \star}\right)^2}{\det {\sf P}^{(2n - 2) \star}}
\Bigg[\frac{{\sf P}^{(2n) \star}}{\det {\sf P}^{(2n - 2) \star}}\
C_{n-2}\left({\sf P}^{(2n - 2) \star}\right)^\star
{\sf P}^{(2n - 4) \star} C_{n-2}\left({\sf P}^{(2n - 2) \star}\right)^\star
\ \ \ \ \ \nonumber\\[0.3cm]
&&\hspace{1.5cm}\ -\ 
C_{n-2}\left({\sf P}^{(2n - 2) \star}\right)^\star \Bigg]
\end{eqnarray}
Emphasizing the role of the effective propagator 
${\sf P}^{(2n - 2) \star}/{\sf P}^{(2n) \star}$ (cf.\ eq.\
(\ref{NE1})) we can rewrite eq.\ (\ref{NE2})
in the following form.
\begin{eqnarray}
\label{NE3}
{\sf A}^{(4) \prime} &=&
-\ \frac{C_{n-2}\left(
\frac{\displaystyle{\sf P}^{(2n - 2) \star}}{
\displaystyle{\sf P}^{(2n) \star}}\right)^\star
}{\det\left(\frac{\displaystyle{\sf P}^{(2n - 2) \star}}{
\displaystyle{\sf P}^{(2n) \star}}\right)}\ 
\frac{{\sf P}^{(2n - 4) \star}}{{\sf P}^{(2n) \star}}\
\frac{C_{n-2}\left(\displaystyle\frac{{\sf P}^{(2n - 2) \star}}{
\displaystyle{\sf P}^{(2n) \star}}\right)^\star
}{\det\left(\frac{\displaystyle{\sf P}^{(2n - 2) \star}}{
\displaystyle{\sf P}^{(2n) \star}}\right)}\ \ \ \ \ \nonumber \\[0.3cm]
&&\hspace{1.5cm}\ +\ 
\frac{C_{n-2}\left(\frac{\displaystyle{\sf P}^{(2n - 2) \star}}{
\displaystyle{\sf P}^{(2n) \star}}\right)^\star
}{\det\left(\frac{\displaystyle{\sf P}^{(2n - 2) \star}}{
\displaystyle{\sf P}^{(2n) \star}}\right)} 
\end{eqnarray}

Unfortunately, the results obtained in the previous subsections
do not yet admit any reliable analytical (heuristic) 
extrapolation to arbitrary
values of $n$ for further quantities beyond
${\sf A}^{(2) \prime}$, ${\sf A}^{(4) \prime}$. 
For example, to heuristically derive an analogous 
expression for ${\sf A}^{(6) \prime}$ one would have to perform
a calculation for $n=5$ first in order to approach this task.
However, in analogy to the previous subsections
it is still possible to derive one further result
for arbitrary $n$.
Again, writing (cf.\ eqs.\ (\ref{NB10}), (\ref{NC14b}), (\ref{ND18}))
\begin{eqnarray}
\label{NE5}
{\sf P}^{(4) \star\prime}&=&
C_2\left({\sf A}^{(2) \prime}\right)^\star\ -\ 
{\sf A}^{(4) \star\prime}
\end{eqnarray}
we find from eqs.\ (\ref{NE1}), (\ref{NE2})
(cf.\ eqs.\
(\ref{NB11}), (\ref{NC15}), (\ref{ND21}))
\begin{eqnarray}
\label{NE6}
{\sf P}^{(4) \prime} &=&
\frac{\left({\sf P}^{(2n) \star}\right)^3}{
\left(\det {\sf P}^{(2n - 2) \star}\right)^2}\
C_{n-2}\left({\sf P}^{(2n - 2) \star}\right)^\star
{\sf P}^{(2n - 4) \star} C_{n-2}\left({\sf P}^{(2n - 2) \star}\right)^\star\ .
\end{eqnarray}
In analogy to eq.\ (\ref{NE3}), this can equivalently be written as
\begin{eqnarray}
\label{NE7}
{\sf P}^{(4) \prime} &=&
\frac{C_{n-2}\left(
\frac{\displaystyle{\sf P}^{(2n - 2) \star}}{
\displaystyle{\sf P}^{(2n) \star}}\right)^\star
}{\det\left(\frac{\displaystyle{\sf P}^{(2n - 2) \star}}{
\displaystyle{\sf P}^{(2n) \star}}\right)}\ 
\frac{{\sf P}^{(2n - 4) \star}}{{\sf P}^{(2n) \star}}\
\frac{C_{n-2}\left(\displaystyle\frac{{\sf P}^{(2n - 2) \star}}{
\displaystyle{\sf P}^{(2n) \star}}\right)^\star
}{\det\left(\frac{\displaystyle{\sf P}^{(2n - 2) \star}}{
\displaystyle{\sf P}^{(2n) \star}}\right)}\ \ \ \ \ 
\end{eqnarray}
To arrive at eq.\ (\ref{NE6}) we have relied on the following
calculation (cf.\ Appendix A, eqs.\ (\ref{A2}), (\ref{A1c})).
\begin{eqnarray}
\label{NE4}
C_2\left({\sf A}^{(2)\prime}\right)~&=&
\left({\sf P}^{(2n) \star}\right)^2\ 
C_2\left(\left[{\sf P}^{(2n-2) \star}\right]^{-1}\right)\nonumber\\[0.3cm]
&=&\left({\sf P}^{(2n) \star}\right)^2\ 
C_2\left({\sf P}^{(2n-2) \star}\right)^{-1}\ =\
\frac{\left({\sf P}^{(2n) \star}\right)^2}{\det {\sf P}^{(2n-2) \star}}\ 
C_{n-2}\left({\sf P}^{(2n-2) \star}\right)^\star\ \ \ \
\end{eqnarray}

Taking the determinant on both sides of the eqs.\ (\ref{NE1}) and
(\ref{NE7}) yields the relations (cf.\ eqs.\
(\ref{NB12}), (\ref{NC17}), (\ref{ND24}) and
(\ref{NC18}), (\ref{ND25}))
\begin{eqnarray}
\label{NE8}
\det {\sf A}^{(2) \prime} &=&
\frac{\left({\sf P}^{(2n) \star}\right)^n}{\det {\sf P}^{(2n-2) \star}}\ ,
\\[0.3cm]
\label{NE9}
\det {\sf P}^{(4) \star\prime}&=&
\frac{\left({\sf P}^{(2n) \star}\right)^{3{n\choose 2}}}{
\left(\det {\sf P}^{(2n-2) \star}\right)^{2(n-1)}}\
\det {\sf P}^{(2n-4)}\ .
\end{eqnarray}
In deriving eq.\ (\ref{NE9}) we have relied on the 
(Sylvester-Franke) identity (cf.\ Appendix A, eq.\ (\ref{A4}))
\begin{eqnarray}
\label{NE10}
\det C_{n-2}\left({\sf P}^{(2n-2) \star}\right) &=& 
\left(\det {\sf P}^{(2n-2) \star}\right)^{n-1\choose n-3}\ .
\end{eqnarray}

\section{The Grassmann integral equation}

Having obtained in the previous section explicit formulas
for the action map $f$ for the case of the Grassmann algebras
${\cal G}_{2n}$, $n = 2, 3, 4$, we can now concentrate
on the study of certain particular relations between 
$G_0[\{\bar\Psi\},\{\Psi\} ]$ and $G[\{\bar\Psi\},\{\Psi\} ]$.
As explained in the Introduction we are interested in 
the equation ($0<\lambda\in{\bf R}$)
\begin{eqnarray}
\label{Q1}
G[\{\bar\Psi\},\{\Psi\} ]&=&G_0[\{\lambda\bar\Psi\},\{\lambda\Psi\} ]
\ +\ \Delta_f(\lambda)\ .
\end{eqnarray}
$\Delta_f(\lambda)$
is some constant which is allowed to depend on $\lambda$ and
which we choose to obey (In view of eq.\ (\ref{M14a}), we have
the freedom to do so.) 
\begin{eqnarray}
\label{Q2}
\Delta_f(1)&=& 0\ .
\end{eqnarray}
For $\lambda = 1$, eq.\ (\ref{Q1}) is the fixed point equation
for the action map $f$ (cf.\ \cite{scha1}, p.\ 288).
Equation (\ref{Q1}) applied to eq.\ (\ref{M7a}),
the latter reads
($\tilde{C} = \exp[-A^{(0)} -\Delta_f(\lambda)]$)
\begin{eqnarray}
\label{Q3}
{\rm e}^{\displaystyle\ G_0[\{\lambda\bar\Psi\},\{\lambda\Psi\} ]}\ &=&
 \tilde{C}\
\int \prod_{l=1}^n\left(d\chi_l\ d\bar\chi_l\right)
\ \ {\rm e}^{\displaystyle\ 
G_0[\{\bar\chi + \bar\Psi\},\{\chi + \Psi\} ]}\nonumber\\[0.3cm]
&&\hspace{2cm}\times\ {\rm e}^{\displaystyle\ \
\sum_{l=1}^n\ \left(\bar\eta_l \chi_l + 
\bar\chi_l \eta_l \right)}\ \ ,\\[0.3cm]
\label{Q4}
\bar\eta_l \ =\
\frac{\partial G_0[\{\lambda\bar\Psi\},\{\lambda\Psi\} ]}{
\partial\Psi_l}\ &,&\ \ \
\eta_l\ =\ -\ 
\frac{\partial G_0[\{\lambda\bar\Psi\},\{\lambda\Psi\} ]}{
\partial\bar\Psi_l}\ .\ \ \ \ \ \ \ \ \ \
\end{eqnarray}
Clearly, this a Grassmann integral equation for
$G_0[\{\bar\Psi\},\{\Psi\} ]$ (more precisely, a nonlinear
Grassmann integro-differential equation).
In view of eq.\ (\ref{M9}), eq.\ (\ref{Q1}) is equivalent to
\begin{eqnarray}
\label{Q5}
A^{(0) \prime}&=&A^{(0)}\ +\ \Delta_f(\lambda)\ ,\\[0.3cm]
\label{Q6}
{\sf A}^{(2k) \prime}&=&\lambda^{2k}\ {\sf A}^{(2k)}\ ,\ \ k>0\ .
\end{eqnarray}
Eq.\ (\ref{Q6}) represents a coupled system of nonlinear matrix
equations.
We are now going to solve eq.\ (\ref{Q3}) (i.e., eq.\ (\ref{Q1}))
for $n = 2, 3, 4$ by solving eq.\ (\ref{Q6}).\\

\subsection{The case $n = 2$}

Applying eq.\ (\ref{Q6}) for $k=1$ to eq.\ (\ref{NB8}),
we find
\begin{eqnarray}
\label{QB1}
{\sf P}^{(4) \star}&=&\lambda^2\ \det{\sf A}^{(2)}\ .
\end{eqnarray}
Eq.\ (\ref{NB3}) then immediately yields
\begin{eqnarray}
\label{QB2}
A_{1 2,1 2}^{(4)}&=&\left(1-\lambda^2 \right)\ \det{\sf A}^{(2)}\ .
\end{eqnarray}
${\sf A}^{(2)}$ remains an arbitrary matrix with
$\det{\sf A}^{(2)}\not= 0$.
To determine $A^{(0)}$ imagine that
the action $G_0[\{\bar\Psi\},\{\Psi\} ]$ specified by eq.\ 
(\ref{QB2}) would have been induced  by some action
$G_{-1}[\{\bar\Psi\},\{\Psi\} ] = 
G_0[\{\lambda^{-1}\bar\Psi\},\{\lambda^{-1}\Psi\} ]$ 
(by means of 
eq.\ (\ref{M7a}) -- replacing $G$ by $G_0$ and $G_0$ by
$G_{-1}$, respectively) with the partition function
$P\left(G_{-1}\right) = 
\lambda^{-2}\det{\sf A}^{(2)}$ (cf.\ eq.\ (\ref{NB3})).
Then (cf.\ footnote \ref{footn3} of Subsect.\ 2.2)
\begin{eqnarray}
\label{QB3}
A^{(0)}&=& \ln P\left(G_{-1}\right)\ =\ \ln\det{\sf A}^{(2)}\ -\ 2 \ln\lambda
\end{eqnarray}
and, consequently,
\begin{eqnarray}
\label{QB4}
\Delta_f(\lambda)&=&4\ \ln\lambda\ .
\end{eqnarray}
From the above considerations we see that for $n=2$ eq.\ (\ref{Q1}) 
has always a solution for any value of $\lambda$ ($0<\lambda\in{\bf R}$).
For $\lambda = 1$ the solution corresponds to a Gaussian integral

while for $\lambda \not= 1$ it corresponds to some non-Gaussian integral
(cf.\ eq.\ (\ref{Q3})).
Consequently, for any even value of $n$ eq.\ (\ref{Q1})
has always a solution for any value of $\lambda$ ($0<\lambda\in{\bf R}$).
This follows from the fact that these solutions can be constructed
as a sum of $n=2$ solutions with a common value of $\lambda$.\\

\subsection{The case $n = 3$}

Applying eq.\ (\ref{Q6}) for $k=1$ to eq.\ (\ref{NC11}),
we find
\begin{eqnarray}
\label{QC1}
{\sf P}^{(6) \star}\ {\bf 1}_3&=&\lambda^2\ {\sf P}^{(4) \star} {\sf A}^{(2)}
\ =\ \lambda^2\ {\sf A}^{(2)} {\sf P}^{(4) \star}\\[0.3cm]
&=&\lambda^2\ \left[ \det{\sf A}^{(2)}\ {\bf 1}_3\ -\ 
{\sf A}^{(4) \star} {\sf A}^{(2)}\right]\nonumber\\[0.3cm]
&=& \lambda^2\ \left[ \det{\sf A}^{(2)}\ {\bf 1}_3\ -\ 
{\sf A}^{(2)} {\sf A}^{(4) \star}\right]\ .
\end{eqnarray}
Furthermore, combining eqs.\ (\ref{NC17}), (\ref{NC18}),
(\ref{NC14b}), (\ref{Q6}) we obtain the relations
\begin{eqnarray}
\label{QC2}
\lambda^6\ \det {\sf P}^{(4) \star} \det {\sf A}^{(2)} &=&
\left({\sf P}^{(6) \star}\right)^3\ ,\\[0.3cm]
\label{QC3}
\lambda^{12}\ 
\frac{\left(\det {\sf P}^{(4) \star}\right)^5}{\det {\sf A}^{(2)}} &=&
\left({\sf P}^{(6) \star}\right)^9\ .
\end{eqnarray}
From these two equations we can conclude that
\begin{eqnarray}
\label{QC4}
\det {\sf P}^{(4) \star}&=&\pm\lambda^3\ \left(\det {\sf A}^{(2)}\right)^2\ ,
\\[0.3cm]
\label{QC5}
{\sf P}^{(6) \star}&=&\pm\lambda^3\ \det {\sf A}^{(2)}\ .
\end{eqnarray}
Taking into account the above equations, from eq.\ (\ref{NC12})
we find then
\begin{eqnarray}
\label{QC6}
{\sf A}^{(4) \star}&=&-\ \left(1\mp \frac{1}{\lambda}\right)\ 
{\sf P}^{(4) \star}\ .
\end{eqnarray}
By virtue of eq.\ (\ref{NC3c}) this entails
\begin{eqnarray}
\label{QC7}
{\sf A}^{(4) \star}&=&(1\mp\lambda)\
{\rm adj}\; {\sf A}^{(2)}\ ,\\[0.3cm]
\label{QC8}
{\sf P}^{(4) \star}&=&\pm\lambda\ {\rm adj}\; {\sf A}^{(2)}\ .
\end{eqnarray}
One easily sees that eq.\ (\ref{QC8}) is in line with the 
result (\ref{QC4}). Finally, applying eqs.\ (\ref{Q6}),
(\ref{QC1}), (\ref{QC4}), (\ref{QC5}) to eq.\ (\ref{NC13})
we calculate $A_{1 2 3,1 2 3}^{(6)}$. It reads
\begin{eqnarray}
\label{QC9}
A_{1 2 3,1 2 3}^{(6)}&=&(\lambda\mp 1)^2 (\pm\lambda- 4)\ 
\det {\sf A}^{(2)}\ .
\end{eqnarray}
Applying the same procedure to eq.\ (\ref{NC16}), 
we find the consistency equation
\begin{eqnarray}
\label{QC11}
(\lambda\mp 1)^3&=&0
\end{eqnarray}
which has only one solution, namely $\lambda = 1$ (choose the upper sign).
This solution is just the elementary one which 
corresponds to a Gaussian integral (cf.\ eq.\ (\ref{Q3})).\\

\subsection{The case $n = 4$}

Applying eq.\ (\ref{Q6}) for $k=1$ to eq.\ (\ref{ND11}),
we find
\begin{eqnarray}
\label{QD1}
{\sf P}^{(8) \star}\ {\bf 1}_4&=&\lambda^2\ {\sf P}^{(6) \star} {\sf A}^{(2)}
\ =\ \lambda^2\ {\sf A}^{(2)} {\sf P}^{(6) \star}\\[0.3cm]
&=&\lambda^2\ \left[ \det{\sf A}^{(2)}\ {\bf 1}_4\ -\ 
{\sf F}_a\left({\bf 1}_4,C_2\left({\sf A}^{(2)}\right)^\star
{\sf A}^{(4) \star}\right)\ -\
{\sf A}^{(6) \star} {\sf A}^{(2)}\right]\ \ \ \ \ \nonumber\\[0.3cm]
&=&\lambda^2\ \left[ \det{\sf A}^{(2)}\ {\bf 1}_4\ -\ 
{\sf F}_a\left({\bf 1}_4,{\sf A}^{(4) \star}
C_2\left({\sf A}^{(2)}\right)^\star\right)\ -\
{\sf A}^{(2)} {\sf A}^{(6) \star}\right]\ .\ \ \ \ 
\end{eqnarray}
Furthermore, combining eqs.\ (\ref{ND24}), (\ref{ND25}),
(\ref{ND18}), (\ref{Q6}) we obtain the relations
\begin{eqnarray}
\label{QD2}
\lambda^8\ \det {\sf P}^{(6) \star} \det {\sf A}^{(2)} &=&
\left({\sf P}^{(8) \star}\right)^4\ ,\\[0.3cm]
\label{QD3}
\lambda^{24}
\left(\det {\sf P}^{(6) \star}\right)^6&=&
\left({\sf P}^{(8) \star}\right)^{18}\ .
\end{eqnarray}
From these two equations we can conclude that
\begin{eqnarray}
\label{QD4}
\det {\sf P}^{(6) \star}&=&\lambda^8\ \left(\det {\sf A}^{(2)}\right)^3\ ,
\\[0.3cm]
\label{QD5}
{\sf P}^{(8) \star}&=&\pm\lambda^4\ \det {\sf A}^{(2)}\ .
\end{eqnarray}
We can now apply eq.\ (\ref{Q6}) to the eqs.\ (\ref{ND12})
and (\ref{ND21}). Taking into account eqs.\ (\ref{ND3h}),
(\ref{ND18}), we can derive from these two equations the following
compound matrix equation.
\begin{eqnarray}
\label{QD6}
C_2\left({\sf P}^{(6) \star} {\sf A}^{(2)}\right)&=&
\left(\lambda^2 \det {\sf A}^{(2)}\right)^2\ {\bf 1}_6
\end{eqnarray}
Its solution reads (cf.\ \cite{marc2}, Sect.\ 3, p.\ 149,
eq.\ (11))
\begin{eqnarray}
\label{QD7}
{\sf P}^{(6) \star}&=&\pm\lambda^2\ {\rm adj}\; {\sf A}^{(2)}\ .
\end{eqnarray}
Eq.\ (\ref{QD7}) is in line with eq.\ (\ref{QD4}) (The signs on
the r.h.s.\ are fixed by making reference to eqs.\ (\ref{QD1}),
(\ref{QD5}).).
We can now take into account eq.\ (\ref{QD7}) in considering
eq.\ (\ref{ND21}) further. Eq.\ (\ref{ND21}) then yields the 
following matrix equation.
\begin{eqnarray}
\label{QD8}
{\sf P}^{(4)} C_2\left({\sf A}^{(2)}\right)^\star&=&\pm\
C_2\left({\sf A}^{(2)}\right) {\sf P}^{(4) \star}
\end{eqnarray}
By virtue of eq.\ (\ref{ND3h}), eq.\ (\ref{QD8}) can equivalently
be written as
\begin{eqnarray}
\label{QD9}
{\sf A}^{(4)} C_2\left({\sf A}^{(2)}\right)^\star&=&\pm\
C_2\left({\sf A}^{(2)}\right) {\sf A}^{(4) \star}\ .
\end{eqnarray}
We will not study here the complete set of solutions of eq.\ 
(\ref{QD9}). This would need to be the subject of a study in its
own. Here, it suffices to mention that for the Ansatz ($\alpha$ is some 
arbitrary constant, ${\sf B}$ some $4\times 4$ matrix)
\begin{eqnarray}
\label{QD10}
{\sf A}^{(4)}&=&\alpha\
C_2\left({\sf B}\right)^\star\ .
\end{eqnarray}
eq.\ (\ref{QD9}) reads
\begin{eqnarray}
\label{QD11}
C_2\left({\sf A}^{(2)} {\sf B}\right)^\star&=&\pm\
C_2\left({\sf A}^{(2)} {\sf B}\right)\ .
\end{eqnarray}
For the upper sign, this is exactly the type of compound matrix equation
studied in \cite{marc2}. Of course, eq.\ (\ref{QD9}) has solutions
which correspond to two $n=2$ solutions (with a common value of
$\lambda$) discussed at the end of
Subsect.\ 3.1 \footnote{Then, ${\sf A}^{(2)}$
has a $2\times2$ block matrix structure and ${\sf A}^{(4)}$
has a diagonal matrix structure with 
${\sf A}^{(4)} = 
{\rm diag}\left(A^{(4)}_{1 2, 1 2},0,0,0,0,A^{(4)}_{3 4, 3 4}\right)$.}.
Here, we want to go beyond these solutions.\\

For the present purpose, we consider in eq.\ (\ref{QD9}) only 
the upper sign on the r.h.s.\ and study the Ansatz 
($\kappa\in {\bf R}$)
\begin{eqnarray}
\label{QD12}
{\sf P}^{(4)}&=&\kappa\ C_2\left({\sf A}^{(2)}\right)\ ,\\[0.3cm]
\label{QD13}
{\sf A}^{(4)}&=&(1-\kappa)\ C_2\left({\sf A}^{(2)}\right)\ ,
\end{eqnarray}
which is a special version of eq.\ (\ref{QD10}).
Inserting this Ansatz into eq.\ (\ref{ND14}) and
taking into account eqs.\ (\ref{Q6}), (\ref{QD5}),  (\ref{QD7}),
we find 
\begin{eqnarray}
\label{QD14}
{\sf A}^{(6) \star}&=&  \left(\lambda^2 - 6 \kappa^2
+ 9 \kappa  - 4\right)\
{\rm adj}\; {\sf A}^{(2)}\ .
\end{eqnarray}
Applying the same procedure to eq.\ (\ref{ND22a}), we obtain
the following consistency condition.
\begin{eqnarray}
\label{QD15}
\lambda^2 - 3 \kappa^2 + 3 \kappa - 1 &=&
\lambda^2 - 3 \kappa (\kappa - 1) - 1\ =\ 0
\end{eqnarray}
Furthermore, applying the Ansatz (\ref{QD12}) to eq.\ (\ref{ND17}) and
taking into account eqs.\ (\ref{Q6}), (\ref{QD5}),  (\ref{QD7})
yields
\begin{eqnarray}
\label{QD16}
A_{1 2 3 4, 1 2 3 4}^{(8)}&=&(\lambda^4 + 20 \lambda^2  
- 24 \lambda^2\kappa + 72 \kappa^3
- 147 \kappa^2
+ 108 \kappa - 30)\ \det{\sf A}^{(2)}\ .\ \ \ \
\end{eqnarray}
Again, subjecting eq.\ (\ref{ND23}) to the same procedure we obtain
yet another consistency condition.
\begin{eqnarray}
\label{QD17}
2 \lambda^2 - 3 \lambda^2 \kappa + 9 \kappa^3 - 15 \kappa^2 
+ 9 \kappa - 2&=&\nonumber\\[0.3cm]
(2 - 3 \kappa) 
\left[\lambda^2 - 3 \kappa (\kappa - 1) - 1 \right]&=&
0
\end{eqnarray}
Obviously, this equation is fulfilled if $\lambda$, $\kappa$
obey eq.\ (\ref{QD15}). Consequently, we can restrict 
our attention to solutions of eq.\ (\ref{QD15}).\\

From eq.\ (\ref{QD15}) we conclude that the Ansatz (\ref{QD12})
provides us with solutions of eq.\ (\ref{Q1}) for any value
of $\lambda\ge 1/2$ (if $\kappa$ assumes real values only).
Of particular interest to us are solutions of eq.\ (\ref{QD15})
for $\lambda = 1$ (see \cite{prok1,scha1}). 
In this case, eq.\ (\ref{QD15}) reads
\begin{eqnarray}
\label{QD18}
\kappa\ (\kappa -1)&=&0\ .
\end{eqnarray}
Clearly, this equation has two solutions:
\begin{eqnarray}
\label{QD19}
\kappa_I&=&1\ ,\\[0.3cm]
\kappa_{II}&=&0\ .
\end{eqnarray}
The corresponding expressions for the action $G_0 = G$ then read
as follows.
\begin{eqnarray}
\label{QD20a}
G_{0I}[\{\bar\Psi\},\{\Psi\} ]&=&G_{0I}(G_q)\nonumber\\[0.3cm]
&=&\ln\det{\sf A}^{(2)}\ + \ G_q\\[0.3cm]
\label{QD20b}
G_{0II}[\{\bar\Psi\},\{\Psi\} ]&=&G_{0II}(G_q)\nonumber\\[0.3cm]
&=&\ln\det{\sf A}^{(2)}\ + \ G_q\ -\ \frac{1}{2}\ G_q^2\ +\ 
\frac{1}{2}\ G_q^3\ -\ \frac{3}{8}\ G_q^4
\\[0.3cm]
\label{QD21}
G_q&=&G_q[\{\bar\Psi\},\{\Psi\} ]\ =\ 
\sum_{l,m=1}^4 A_{l,m}^{(2)} \bar\Psi_l \Psi_m\ =\ 
\bar\Psi {\sf A}^{(2)} \Psi
\end{eqnarray}
As one can see from eq.\ (\ref{Q3}) $G_{0I}$ corresponds to a 
Gaussian integral while $G_{0II}$ corresponds to some non-Gaussian
integral. While it is well-known that for the action $G_0=G_{0I}$
the equation $G = G_0$ applies it is indeed a remarkable 
fact that the same is true for $G_0=G_{0II}$. However, this is not
yet the end of remarkable features of these actions.
It is also instructive to work out for $\kappa_I = 1$ and
$\kappa_{II} = 0$ the corresponding expressions
for $W[\{\bar\eta\},\{\eta\} ]$ on the basis of eq.\ (\ref{ND5}).
\begin{eqnarray}
\label{QD22a}
W_I[\{\bar\eta\},\{\eta\} ]&=&W_I(W_q)\ =\ G_{0I}(W_q) \nonumber\\[0.3cm]
&=&\ln\det{\sf A}^{(2)}\ + \ W_q\\[0.3cm]
\label{QD22b}
W_{II}[\{\bar\eta\},\{\eta\} ]&=&W_{II}(W_q)\ =\ 
G_{0II}(W_q)\nonumber\\[0.3cm]
&=&\ln\det{\sf A}^{(2)}\ + \  
W_q\ -\ \frac{1}{2}\ W_q^2\ +\ \frac{1}{2}\ W_q^3\ -\ 
\frac{3}{8}\ W_q^4
\\[0.3cm]
\label{QD23}
W_q&=&W_q[\{\bar\eta\},\{\eta\} ]\ =\ 
-\ \bar\eta \left[{\sf A}^{(2)}\right]^{-1} \eta
\end{eqnarray}
Again, while the relation $W_I = G_{0I}$ is well-known
in the present context 
the equality $W_{II}= G_{0II}$ comes as a complete
surprise and one can only wonder which general principle is 
manifesting here itself. We will explore this issue
in the next subsection.\\

\subsection{Further analysis}

We can characterize the solutions $G_{0I}$, $G_{0II}$ 
of the equation (\ref{Q1}) found for $n=4$, $\lambda = 1$,
in the previous subsection
by two properties which may be of general significance.
The first one is related to the identity $W = G_0$
(eqs.\ (\ref{QD22a}), (\ref{QD22b})). 
One immediately recognizes that for 
\begin{eqnarray}
\label{QE1}
\left[{\sf A}^{(2)}\right]^2&=&-{\bf 1}_4
\end{eqnarray}
$\exp G_0 = \exp G_{0I} (= \exp G = Z)$ and $\exp G_0 =\exp G_{0II}$ are 
{\it self-reciprocal} Grassmann functions
(Of course, this is a well-known property of $\exp G_{0I}$.):
\begin{eqnarray}
\label{QE2}
\int \prod_{l=1}^4\left(d\chi_l\ d\bar\chi_l\right)
\ \ {\rm e}^{\displaystyle\ 
G_0[\{\bar\chi\},\{\chi\} ]\ +\ 
\bar\eta \chi\ +\ \bar\chi \eta}
&=&{\rm e}^{\displaystyle\ G_0[\{\bar\eta\},\{\eta\} ]}\end{eqnarray}
($\det{\sf A}^{(2)}= 1$, cf.\ eq.\ (\ref{QE1}))\footnote{We disregard
here the generalization to the case $\det{\sf A}^{(2)}= -1$.}, 
i.e., they are eigenfunctions to 
the Fourier-Laplace transformation\footnote{If we would re-define
the Fourier-Laplace transformation in a Grassmann algebra
by replacing on the l.h.s.\ of eq.\ (\ref{QE2}) $\bar\eta$,
$\eta$ by $i\bar\eta$, $i\eta$, eq.\ (\ref{QE1}) would of course
read $\left[{\sf A}^{(2)}\right]^2 = {\bf 1}_4$.} 
to the eigenvalue 1.
The term `self-reciprocal function' is taken from real (complex) analysis
where it also denotes eigenfunctions of some integral
transformation, in particular, the Fourier transformation\footnote{
Cf., e.g., \cite{hard1}, \cite{titc1}, Chap.\ IX, p.\ 245. 
For an early account of the history
of self-reciprocal functions see \cite{mehr1}. 
Eigenfunctions to the eigenvalue $-1$ are often called
`skew-reciprocal'. In optics, following
a paper by Caola \cite{caol} (who seems to not have been aware of the history
of the subject in mathematics) in recent years self-reciprocal
functions are often referred to as `self-Fourier functions' (in 
the context of the Fourier transformation). Incidentally, 
also note the comment
in \cite{bane} on the earlier optics literature on the subject.}.\\

The second property of the solutions $G_{0I}$, $G_{0II}$ 
is related to the identity $G = W$ (apply the fixed point 
condition $G = G_0$ to the eqs.\ (\ref{QD22a}), (\ref{QD22b})). 
Taking into account eqs.\
(\ref{QD20a})-(\ref{QD23}), eqs.\  (\ref{M7c0}), (\ref{M7b})
tell us that
\begin{eqnarray}
\label{QE3}
G(G_q) &=& G(W_q)
- \sum_{l=1}^4\ \left(\bar\eta_l \Psi_l + 
\bar\Psi_l \eta_l \right)\ ,\\[0.3cm]
\label{QE4}
\bar\eta_l &=&\ \ \ 
\frac{\partial G(G_q)}{\partial\Psi_l}\ =\ 
-\ G^{\;\prime}(G_q)\ \left(\bar\Psi {\sf A}^{(2)}\right)_l\ ,\\[0.3cm]
\label{QE5}
\eta_l&=& -\ 
\frac{\partial G(G_q)}{\partial\bar\Psi_l}\ =\ 
-\ G^{\;\prime}(G_q)\ \left({\sf A}^{(2)} \Psi\right)_l\ .
\end{eqnarray}
Here, 
\begin{eqnarray}
\label{QE6}
G^{\;\prime}(G_q)&=&\frac{\partial G(G_q)}{\partial G_q}
\end{eqnarray}
where $G_q$ is treated as a formal parameter for the moment.
In view of eqs.\ (\ref{QE4}), (\ref{QE5}) it holds
\begin{eqnarray}
\label{QE7}
W_q&=&- G_q\ \left[G^{\;\prime}(G_q)\right]^2 \ .
\end{eqnarray}
Taking into account the eqs.\ (\ref{QE4}), (\ref{QE5}), (\ref{QE7}),
eq.\ (\ref{QE3}) can be written as
\begin{eqnarray}
\label{QE8}
G(s)&=&G\left(-s \left[G^{\;\prime}(s)\right]^2\right)\ +\
2 s\ G^{\;\prime}(s)\ ,\ \ s\ =\ G_q\ .
\end{eqnarray}
Eq.\ (\ref{QE8}) is of a very general nature. Its shape does not
depend on the value of $n$. Its derivation depends on 
the fact only that $G$, $W$ are functions of $G_q$, $W_q$, respectively,
and that the identity $G = W$ holds. As we demonstrate in 
Appendix C, eq.\ (\ref{QE8}) can also be derived under analogous
conditions starting from a (Euclidean space-time) version
of eqs.\ (\ref{M1a})-(\ref{M2}) for a scalar field 
on a finite lattice. Consequently, until further notice we disregard
the fact that $s$ is a bilinear in the Grassmann algebra generators
and simply understand eq.\ (\ref{QE8}) as an equation for
a function $G = G(s)$. We will now analyze eq.\ (\ref{QE8}) further.\\

Eq.\ (\ref{QE8}) appears to be somewhat involved but it
can be simplified the following way.
We can differentiate both sides of eq.\ (\ref{QE8}) with
respect to $s$. The resulting equation can be transformed to
read
\begin{eqnarray}
\label{QE9}
\left\{ 2 s\ G^{\;\prime\prime}(s) + G^{\;\prime}(s)\right\}
\left\{ 1 - G^{\;\prime}(s)\
G^{\;\prime}\left(- s \left[G^{\;\prime}(s)\right]^2\right)\right\}
&=&0\ .
\end{eqnarray}
Eq.\ (\ref{QE9}) is being obeyed if either one of the two 
following equations of very different mathematical nature
is respected.
\begin{eqnarray}
\label{QE10}
2 s\ G^{\;\prime\prime}(s) + G^{\;\prime}(s)&=&0\\[0.3cm]
\label{QE11}
G^{\;\prime}(s)\
G^{\;\prime}\left(- s\ \left[G^{\;\prime}(s)\right]^2\right)
&=&1
\end{eqnarray}
The solution of the linear differential equation (\ref{QE10}) reads
\begin{eqnarray}
\label{QE12}
G^{\;\prime}(s)&\sim&{\rm e}^{\displaystyle\ - \sqrt{s}}
\end{eqnarray}
entailing
\begin{eqnarray}
\label{QE13}
G(s)&\sim&\left(1 + \sqrt{s}\right)\ {\rm e}^{\displaystyle\ - \sqrt{s}}\ .
\end{eqnarray}
As $G(s)$ depends on $\sqrt{s}$ this solution is of no relevance
in the context of Grassmann algebras. To see this note that the function
$G(s)$ contains odd powers of $\sqrt{s}$ in its (Taylor) expansion
(in terms of $t = \sqrt{s}$) around $s = 0$.
If $s$ is being interpreted
as a bilinear form in the generators of the Grassmann algebra
these terms have no interpretation within the 
Grassmann algebra framework\footnote{Incidentally, in 
considering the bosonic (real/complex analysis) analogue (eq.\ (\ref{T1}))
of the Grassmann integral equation (\ref{Q3}) 
eq.\ (\ref{QE13}) can also be disregarded as it does not lead
to any convergent integral.}. Consequently, in
the following we can concentrate our attention onto 
the nonlinear functional equation (\ref{QE11}).\\

To gain further insight it turns out to be convenient now
to define the following functions (The definition in eq.\ (\ref{QE15})
could equally well read $d(t) = -i\ b(t)$.).
\begin{eqnarray}
\label{QE14}
b(t)&=&t\ G^{\;\prime}(t^2)\ =\ \frac{1}{2}\ 
\frac{\partial}{\partial t} G(t^2)\\[0.3cm]
\label{QE15}
d(t)&=& i\ b(t)
\end{eqnarray}
Then, having multiplied both sides by $-\sqrt{s}$ 
eq.\ (\ref{QE11}) can be written as
($t = \sqrt{s}$)
\begin{eqnarray}
\label{QE16}
d^2(t)&=&d\left(d(t)\right)\ =\ - t\ .
\end{eqnarray}
This is an iterative functional equation: the function $d(t)$ is
the (second) iterative root of $-{\bf 1}$ (for a review of 
iterative
functional equations see \cite{kucz2}, in particular Chap.\ 11, 
p.\ 421, \cite{kucz1}, in particular Chap.\ XV, p.\ 288,
also see \cite{targ}, Chap.\ 2, p.\ 36). The functional equation
(\ref{QE16}) has been studied by Massera and Petracca \cite{mass}
who have pointed out its relation to the equivalent functional
equation 
\begin{eqnarray}
\label{QE16b}
h\left(h(x)\right)&=& \frac{1}{x}
\end{eqnarray}
(Define the involution $q(x) = (1-x)/(1+x)$. If $h(x)$ is a
solution of eq.\ (\ref{QE16b}) the function $q\circ h\circ q$
is a solution of eq.\ (\ref{QE16})). This functional equation 
characterizes functions $h$ for which $h^{-1}=1/h$ 
(note in this context
\cite{eule}-\cite{chen}, in particular \cite{chen}, p.\ 712).
Eq.\ (\ref{QE16}) has also been studied for real functions
in \cite{pely}, Chap.\ II, \S 5, p.\ 54, and in \cite{falc}-\cite{mcca3}.
Of course, in view of eq.\ (\ref{QE15})
in general we are concerned with complex solutions of eq.\ (\ref{QE16}).\\

If the function $G^{\;\prime} (s)$ has a definite symmetry under 
$s\longrightarrow -s$ eq.\ (\ref{QE16}) can be simplified
to some extent (getting rid of the imaginary unit $i$ present
in eq.\ (\ref{QE15})). If $G^{\;\prime} (s)$ is an even function
(i.e., up to some constant $G(s)$ is odd) 
eq.\ (\ref{QE16}) can be written as
\begin{eqnarray}
\label{QE17}
b^2(t)&=&b\left(b(t)\right)\ =\ t\ .
\end{eqnarray}
This iterative functional equation is 
a special case of the functional equation
$b^k(t) = t$ which is being called 
the {\it Babbage equation} (It has been studied
first by Charles Babbage \cite{babb1a}-\cite{camp}.
See \cite{kucz1}, Chap.\ XV, \S 1, p.\ 288,
\cite{kucz2}, Sect.\ 11.6, p.\ 450, for more information
and references, recent references not referred to in \cite{kucz1},
\cite{kucz2} are \cite{mcca4}, \cite{lait}.).
Solutions $b(t)$ of eq.\ (\ref{QE17}) 
(i.e., solutions of the Babbage equation for $k=2$) are called
{\it involutory functions} {\it ((second) iterative roots of 
unity/identity}, {\it periodic functions/maps)}. 
If, for example, the function $G(s)$ stands in correspondence
to a Gaussian integral (cf.\ eq.\ (\ref{Q3})), 
$G(s) = s$ and, consequently,
\begin{eqnarray}
\label{QE18}
b(t)&=&t\ .
\end{eqnarray}
This is the most elementary involutory function one can think of.
Note, that the set of solutions of eq.\ (\ref{QE17}) is very large
as this set is equivalent to the set of even function (see, e.g.,
\cite{acze}, \cite{schw}, \cite{kucz2}, p.\ 451).
If $G^{\;\prime} (s)$ is an odd function
(i.e., $G(s)$ is even) eq.\ (\ref{QE16}) can be written as
\begin{eqnarray}
\label{QE19}
b^2(t)&=&b\left(b(t)\right)\ =\ - t\ .
\end{eqnarray}
However, this case is not very interesting as real functions
solving eq.\ (\ref{QE19}) are necessarily discontinuous 
(\cite{kucz1}, Chap.\ XV, \S 4, p. 299, \cite{falc},
\cite{eule}-\cite{chen}, \cite{mcca1}, \cite{kucz2},
Subsect.\ 11.2B, p.\ 425).\\

The above consideration can be applied to the solutions of the 
Grassmann integral equation found in Subsect.\ 3.3.
Eq.\ (\ref{QD20a}) is of course being described by eq.\ (\ref{QE18})
($b_I(t) = t$).
From eq.\ (\ref{QD20b}) we recognize that the function 
$G_{II}(s)$ does not have a definite symmetry under 
$s\longrightarrow -s$. We find
\begin{eqnarray}
\label{QE20}
b_{II}(t)&=&t\ \left(1 - t^2 +\frac{3}{2}\ t^4 - \frac{3}{2}\ t^6\right)
\end{eqnarray}
and one can check that the corresponding function 
$d_{II}(t) = i b_{II}(t)$
indeed fulfills eq.\ (\ref{QE16}) at order $t^7$ (Going through the
above arguments one can convince oneself that this is the 
appropriate order in $t$
one has to take into account for the Grassmann algebra ${\cal G}_8$.
Order $t^7$ corresponds to order $s^3$ in eq.\ (\ref{QE11}).).\\

\newpage
\section{Discussion and conclusions}

While most of the explicit expressions obtained in the present paper
for the Grassmann algebras ${\cal G}_{2n}$, $n= 2,3,4$, have been
obtained here for the first time, some of them can be compared to
results derived earlier by other authors. The point is that
partition functions for specific (finite-dimensional)
fermionic systems have been obtained by
a number of authors and some of these results can be used for 
direct comparison with the present results. For example, our expressions
(\ref{NB3}), (\ref{NC4b}), (\ref{ND4b}), can be seen to agree with
eq.\ (8), p.\ 694, of \cite{ulla1}. Furthermore, our eq.\ (\ref{NB3})
is in line with eq.\ (13), p.\ 1298, of \cite{deso1}, the same
applies to our eq.\ (\ref{NC4b}) and its counterpart, eq.\ (14),
p.\ 1298, \cite{deso1}. Also eq.\ (16), p.\ 1298, \cite{deso1} (for 
$n=3$, $l=3$ and $n=4$, $l=2,3,4$) gives the same results as
our eqs.\ (\ref{NC4b}), (\ref{ND4b}). And finally, our eq.\ 
(\ref{ND4b}) agrees with eq.\ (10), p.\ 1083, of \cite{gran1} (for
$N=4$).\\

Our consideration of the action map $f$ in the present paper has
been motivated by the formalism of (lattice) quantum field theory.
However, the consideration of certain modifications of the map
$f$ might also be of some interest from various points of view.
Let us consider a special set of modifications which 
can be described by replacing the eqs.\ (\ref{M7b}) by the 
equations 
\begin{eqnarray}
\label{F1}
\bar\eta_l &=&
\frac{\partial 
\tilde{G}[\{\bar\Psi\},\{\Psi\} ]}{\partial\Psi_l}\ \ ,\ \ \ \
\eta_l\ =\ -\ 
\frac{\partial \tilde{G}[\{\bar\Psi\},\{\Psi\} ]}{\partial\bar\Psi_l}
\end{eqnarray}
($G$ is replaced by $\tilde{G}$). 
For example, if one is just interested in the fixed point condition
for the action map $f$ (i.e., in the eq.\ (\ref{Q1}) for $\lambda=1$)
it might make sense to consider instead of the action map $f$
a different map $\tilde{f}$ (described by the eqs.\ (\ref{M7b}), (\ref{F1}),
respectively)
having the same set of fixed points 
but which is algebraically or numerically easier to study. One such 
modification consists in choosing $\tilde{G}=G_0$
(cf.\ \cite{scha1}, p.\ 291, eq.\ (2.9)).
The implicit representation of the map $f$ given in eqs.\ 
(\ref{M7a}), (\ref{M7b}) would then turn into an explicit representation
of the map $\tilde{f}$ which might be easier to handle in some
respect. As an aside in this context, we mention that for this
map $\tilde{f}$ the equations (\ref{NB6a}), (\ref{NC6}), (\ref{ND6})
(replace ${\sf A}^{(2) \prime}$ on the r.h.s.\ by ${\sf A}^{(2)}$)
exhibit a \underline{formal} similarity to the main equation for the Schulz
iteration scheme for the calculation of the inverse of a matrix
(see eq.\ (7), p.\ 58, in \cite{schu}\footnote{Also see in this 
respect, e.g., \cite{fraz}, Chap.\ IV, Part I, Art.\ 4.11, p.\ 120,
\cite{hote}, Sect.\ 7, p.\ 14, \cite{bode}, Part III.B, p.\ 227.
For further references see, e.g., \cite{pier}.}).
The similarity, however, is only formal as in general the matrix
${\sf P}^{(2n - 2) \star}/{\sf P}^{(2n) \star}$ is not 
invariant under the map $\tilde{f}$ (For the simplest case, $n=2$, 
for example, one can convince oneself of this fact starting from eqs.\
(\ref{NB6a}), (\ref{NB6b}) where one has to omit in this 
case the primes on the r.h.s..).\\

As already mentioned the investigation performed in the
present study within the framework of Grassmann algebras
has been inspired by a problem in quantum field theory which in its
simplest version (within zero-dimensional field theory)
is a problem in real/complex analysis. The standard analysis
analogue of the Grassmann integral equation studied in Chap.\ 3
(for $\lambda = 1$) reads (cf.\ eq.\ (\ref{M5}))
\begin{eqnarray}
\label{T1}
{\rm e}^{\displaystyle\ g(y)}\ &=& C\
\int\limits^{+\infty}_{-\infty} dx\ \ {\rm e}^{\displaystyle\ g(x+y) 
\ -\ g^\prime (y) x}\ .
\end{eqnarray}
This is a nonlinear integro-differential equation for the real 
function $g(x)$. Clearly, the elementary function $g(x) = -a x^2/2$, 
$0< a\in {\bf R}$ ($C = \sqrt{a/(2\pi)}$) solves this equation.
However, the interesting question is if this equation has any
other (non-elementary) solution which stands in correspondence to
a non-Gaussian integral. No qualitative information seems to 
be available in the mathematical literature in this respect.
As pointed out in \cite{prok1}, Sect.\ 4, p.\ 859 (p.\ 475 of the 
English transl.), eq.\ (\ref{T1}) is a very complicated 
equation. The main difficulty in explicitly finding any 
non-elementary solution to it (if it exists at all -- we just assume
this for the time being) consists in the fact that it is
very difficult if not impossible to 
calculate for an arbitrary function $\exp g(x)$
its Fourier (or Laplace) transform explicitly. The question now
arises if the analysis in Subsect.\ 3.4 of the solutions 
of the Grassmann integral equation found for $n=4$, $\lambda = 1$,
might help in overcoming this problem. We do not have any final

answer on this but in our view it makes sense to say: perhaps.
The solutions of the Grassmann integral equation found for 
$n=4$, $\lambda = 1$, are characterized by two properties which 
are not related to the anticommuting character of Grassmann
variables. The solutions were related, first, to eigenfunctions 
of the Fourier-Laplace transformation to the eigenvalue 1
(i.e., to self-reciprocal
functions) and, second, to some iterative functional equation.
Now, it seems to be reasonable to assume that also (some) solutions
of eq.\ (\ref{T1}) might be characterized by these two properties.
The two sets of functions obeying one of these two principles are very large
and one might think that the intersection of these two sets
contains also other functions than just the functions given by
$g(x) = -a x^2/2$. The task of solving eq.\ (\ref{T1})
then is equivalent to studying eigenfunctions of the Fourier
transformation to the eigenvalue 1, i.e., self-reciprocal
functions $\exp g(x)$\footnote{For a general discussion of eigenfunctions
and eigenvalues of integral transformations see \cite{doet}, for 
a discussion concerning the Fourier transformation see, e.g., 
\cite{titc1}, Subsect.\ 3.8, p.\ 81. Incidentally and as an
aside, we mention here that the Laplace transformation does not have
any real eigenfunctions to the eigenvalue 1 \cite{hard2}, 
\cite{doet}, \S 4, p.\ 121.}. 
They obey the equation
\begin{eqnarray}
\label{T2}
{\rm e}^{\displaystyle\ g(y)}\ &=&\
\int\limits^{+\infty}_{-\infty} \frac{dx}{\sqrt{2\pi}}\ \ 
{\rm e}^{\displaystyle i y x}\ 
{\rm e}^{\displaystyle\ g(x)}\ .
\end{eqnarray}
The consideration of eigenfunctions of the Fourier transformation
solves the above mentioned problem of finding their Fourier 
transforms at once\footnote{Possibly, it might be appropriate here to
not loose sight of the fact that besides the eigenvalue 1
the Fourier transformation has also three more eigenvalues: -1, $\pm i$.}.
There is a vast mathematical 
literature on self-reciprocal functions (in particular
for the Fourier transformation) but in our context it makes sense
to concentrate on a certain subclass of self-reciprocal functions.
Klauder \cite{klau4}, p.\ 375, \cite{klau2}, Subsect.\ 10.4, p.\ 246,
has pointed out the relevance of
infinitely divisible characteristic functions in a quantum field
theoretic context. This entails in our context
that the self-reciprocal functions
$\exp g(x)$ should be self-reciprocal probability densities (positive 
definite ones, in addition: without zeros - this follows from infinite
divisibility).
The subject of self-reciprocal (positive definite) 
probability densities
has been studied for some time in probability theory
\cite{levy}-\cite{laue1}, \cite{bond1}, Subsect.\ 7.5, p.\ 122, 
\cite{ross1,schl}, \cite{laue2}, Chap.\ 6, p.\ 148, \cite{nosr}
(see \cite{ross1}, \cite{laue2} for some further references).
Of most relevance to the present problem is the work by Teugels \cite{teug} 
who describes explicit methods to construct solutions of eq.\ 
(\ref{T2}) (also note \cite{nosr} in this respect). From the solutions
$\exp g(x)$ of (\ref{T2}) (which are even functions) we define the function 
$G = G(-x^2/2) = g(x)$\ \footnote{Incidentally, it seems worth
mentioning here a theorem of
Marcinkiewicz \cite{marc3}, p.\ 616 (p.\ 467 of the `Collected Papers'),
theorem $2^{\rm bis}$: If the function 
$g(x)$ is a finite polynomial in $x$ its degree cannot exceed
2. Otherwise, the function $\exp g(x)$ cannot be a
characteristic function.
Consequently, in the (non-Gaussian) cases we are interested 
in the function $g(x)$ cannot be a finite polynomial.}. The function
$d(t)$ (eq.\ (\ref{QE15})) associated with it 
then has to obey the functional equation (\ref{QE16}) in order to 
ensure that the function $g(x)$ solves eq.\ (\ref{T1}). In the
case under discussion, the equations (\ref{QE14})-(\ref{QE16})
can be reformulated the following way. Define the functions
\begin{eqnarray}
\label{T3}
\tilde{b}(x)&=&-\ \frac{\partial g(x)}{\partial x}=\
-\ \frac{\partial}{\partial x} G\left(-\frac{x^2}{2}\right)\ =\
x\ G^{\;\prime}\left(-\frac{x^2}{2}\right)\ ,\\[0.3cm]
\label{T4}
\tilde{d}(x)&=& i\ \tilde{b}(x)\ .
\end{eqnarray}
Then, from eq.\ (\ref{QE16}) one can derive the following 
iterative functional equation which has to be fulfilled.
\begin{eqnarray}
\label{T5}
\tilde{d}^2(x)&=&\tilde{d}\left(\tilde{d}(x)\right)\ =\ - x\ .
\end{eqnarray}
As in Subsect.\ 3.4, one can now assume a certain behaviour
of the function $g(x)$. Assuming again that the function 
$G^{\;\prime} (s)$ is an even function
(i.e., up to some constant $G(s)$ is odd)
eq.\ (\ref{T5}) can be written as
\begin{eqnarray}
\label{T6}
\tilde{b}^2(x)&=&\tilde{b}\left(\tilde{b}(x)\right)\ =\ x\ .
\end{eqnarray}
However, this case is not very interesting as it does not 
lead to any non-Gaussian function $\exp g(x)$ \cite{luka}, Theorem 3,
p.\ 117 (Teor.\ Veroyatn.\ Prim.), p.\ 119 (Theor.\ Prob.\ Appl.;
note that Lukacs uses the term `self-reciprocal' in this article 
in a different sense than we do in the present paper.).
Assuming that $G^{\;\prime} (s)$ is an odd function
(i.e., $G(s)$ is even) eq.\ (\ref{T5}) can be written as
\begin{eqnarray}
\label{T7}
\tilde{b}^2(x)&=&\tilde{b}\left(\tilde{b}(x)\right)\ =\ - x\ .
\end{eqnarray}
However, this case is also not very interesting as real functions
solving eq.\ (\ref{T7}) are necessarily discontinuous 
(\cite{kucz1}, Chap.\ XV, \S 4, p. 299, \cite{falc},
\cite{eule}-\cite{chen}, \cite{mcca1}, \cite{kucz2},
Subsect.\ 11.2B, p.\ 425). Consequently, eq.\ (\ref{T5})
cannot sensibly be simplified by the above considerations.
However, the sketched programme still faces another challenge.
At first glance, it is not obvious how to combine the existent
theory of 
self-reciprocal probability densities with the theory of 
iterative functional equations in an operationally effective
way in order to find non-elementary solutions of eq.\ (\ref{T1})
(or its multidimensional generalizations)
which correspond to non-Gaussian integrals. This will have
to be the subject of further research.\\

This discussion has brought us to the end of the present study.
What are its main results? From a mathematical point of view,
the paper introduces a new type of equation which has not been
studied before -- a Grassmann integral equation. The 
concrete equation studied has been shown to be equivalent
to a coupled system of nonlinear matrix equations which can
be solved (Sect.\ 3). 
From the point of view of standard quantum field theory,
the main results of the present article are as follows.
For low-dimensional Grassmann algebras 
the present paper derives explicit expressions for the 
finite-dimensional analogue of the effective action functional
in terms of the data specifying a fairly general Ansatz
for the corresponding analogue of the so-called ``classical''
action functional (Sect.\ 2). This is a model study which 
in some way
can be understood as the fermionic (Grassmann) analogue 
of zero-dimensional field theory and which may turn out
to be useful in several respect.
Moreover, for an arbitrary Grassmann algebra
(related to an arbitrary purely fermionic ``lattice quantum 
field theory'' -- on a finite lattice) on the basis of the 
explicit results obtained for low-dimensional Grassmann algebras
an exact expression for the four-fermion term of the finite
lattice analogue of the effective action
functional is derived in a heuristic manner (Subsect.\ 2.5,
eq.\ (\ref{NE3})). 
From the point of view of the conceptual foundations of 
quantum field theory, the present study demonstrates on the 
basis of a finite-dimensional example that the (Grassmann)
integral equation proposed in \cite{prok1,scha1} can have 
solutions which are equivalent to non-Gaussian integrals (Sect.\ 3).
This certainly will be of interest in various respect.
To illustrate this point let us repeat in compact form some of 
the results found for the Grassmann algebra ${\cal G}_8$
in Subsect.\ 3.3 (These results are specific for this Grassmann
algebra.). Define for an arbitrary invertible
$4\times 4$ matrix ${\sf B}$ ($\det{\sf B}\not=0$) the Grassmann
bilinears
\begin{eqnarray}
\label{T8}
G_q&=&
\sum_{l,m=1}^4 {\sf B}_{lm} \bar\chi_l \chi_m\ =\ 
\bar\chi {\sf B} \chi\ ,\\[0.3cm]
\label{T9}
W_q&=&
-\ \sum_{l,m=1}^4 \left[{\sf B}^{-1}\right]_{lm} \bar\eta_l \eta_m\ =\ 
-\ \bar\eta \left[{\sf B}\right]^{-1} \eta\ .
\end{eqnarray}
Then, the following equation applies.
\begin{eqnarray}
\label{T10}
\hspace{-1cm}\int \prod_{l=1}^4\left(d\chi_l\ d\bar\chi_l\right)
\ \ {\rm e}^{\displaystyle\ (\bar\eta \chi\ +\ \bar\chi \eta)}\ 
\exp\left[ G_q\ -\ \frac{1}{2}\ G_q^2\ +\ 
\frac{1}{2}\ G_q^3\ -\ \frac{3}{8}\ G_q^4\right]&&\nonumber\\[0.3cm]
=\ \det{\sf B}
\ \exp\left[ W_q\ -\ \frac{1}{2}\ W_q^2\ +\ \frac{1}{2}\ W_q^3\ -\ 
\frac{3}{8}\ W_q^4\right]&&
\end{eqnarray}
This should be compared to the well-known, corresponding 
result for a Gaussian integral:
\begin{eqnarray}
\label{T11}
\int \prod_{l=1}^4\left(d\chi_l\ d\bar\chi_l\right)
\ \ {\rm e}^{\displaystyle\ (\bar\eta \chi\ +\ \bar\chi \eta)}\ 
\exp\left[ G_q\right]&=&\det{\sf B}
\ \exp\left[ W_q\right]\ .
\end{eqnarray}
Moreover, in Subsect.\ 3.3 it has been found that the (Grassmann) function
$G_q\ -\ \frac{1}{2}\ G_q^2\ +\ 
\frac{1}{2}\ G_q^3\ -\ \frac{3}{8}\ G_q^4$ is the (first)
Legendre transform of the function 
$W_q\ -\ \frac{1}{2}\ W_q^2\ +\ \frac{1}{2}\ W_q^3\ -\ 
\frac{3}{8}\ W_q^4$ (cf.\ eqs.\ (\ref{QE3})-(\ref{QE5})).
This entails that these functions behave exactly the same way
as the functions $G_q$ and $W_q$. It is clear that any Grassmann
algebra ${\cal G}_{8k}$, $1\le k\in {\bf N}$, supports equations
of the type (\ref{T10}) (simply by multiplying $k$ copies of
eq.\ (\ref{T10})). Given the role that Gaussian
integrals and their properties 
play in quantum field theory, statistical physics and
probability theory it will be interesting to explore 
the implications and applications of the above results in
the future.\\

\vfill

\subsection*{Acknowledgements}

I am grateful to K.\ J.\ Falconer for sending me a reprint of
\cite{falc} and to U.\ Prells for providing me with an 
advance copy of \cite{prel}.\\

\newpage
\setcounter{section}{1}
\section*{Appendix A}
\renewcommand{\theequation}{\mbox{\Alph{section}.\arabic{equation}}}
\setcounter{equation}{0}

Here we collect some formulas for compound matrices\footnote{In the
context of projective geometry, these matrices and their elements
often are referred
to as Pl\"ucker-Grassmann coordinates.}.
Let ${\sf B}$, ${\sf D}$ be $n\times n$ matrices.
The {\it compound matrix} $C_k\left({\sf B}\right)$, 
$0\le k\le n$, is a ${n\choose k}\times {n\choose k}$ 
matrix of all order $k$
minors of the matrix ${\sf B}$. The indices of the compound
matrix entries are given by ordered strings of length $k$.
These strings are composed from the row and 
column indices of the matrix elements of the matrix ${\sf B}$ 
the given minor of the matrix ${\sf B}$ is composed of.
Typically, the entries 
of a compound matrix are ordered lexicographically with respect
to the compound matrix indices (We also apply this convention.).
The {\it supplementary} (or {\it adjugate}) 
{\it compound matrix} $C^{\; n-k}\left({\sf B}\right)$
(sometimes also referred to as the {\it matrix of the $k$th cofactors})
of the matrix ${\sf B}$ is defined by
the equation (cf.\ eq.\ (\ref{M11}))
\begin{eqnarray}
\label{A1d}
C^{\; n-k}\left({\sf B}\right)&=&C_{n-k}\left({\sf B}\right)^\star\ .
\end{eqnarray}
The components of the supplementary compound matrix 
$C^{\; n-k}\left({\sf B}\right)$ can also be defined by means
of the following formula
(here, $l_1 < l_2 <\ldots < l_k$, 
$m_1 < m_2 <\ldots < m_k$; \cite{muir2}, Chap.\ IV, \S 89,
p.\ 75, \cite{vein1}, Chap.\ 3, p.\ 18; also
see our eqs.\ (\ref{M15a})-(\ref{M20})).
\begin{eqnarray}
\label{A1f}
C^{\; n-k}\left({\sf B}\right)_{LM}&=&\frac{\partial}{
\partial {\sf B}_{l_1 m_1}}\ldots
\frac{\partial}{\partial {\sf B}_{l_k m_k}}\ \det{\sf B}
\end{eqnarray}
This comparatively little known definition of (matrices of)
cofactors (supplementary
compound matrices) is essentially due to Jacobi \cite{jaco2}, \S 10,
p.\ 301, p.\ 273 of the `Gesammelte Werke', p.\ 25 of the German transl.\
(also see
the corresponding comment by Muir in \cite{muir1}, Part I, Chap.\ IX,
pp.\ 253-272, in particular pp.\ 262/263).\\

For compound matrices holds (${\bf 1}_r$ is the $r\times r$ unit matrix,
$\alpha$ some constant)
\begin{eqnarray}
\label{A1b}
C_k\left(\alpha {\bf 1}_n\right)&=&\alpha^k\ {\bf 1}_{n\choose k}\ .
\end{eqnarray}
Important relations are given by the 
{\it Binet-Cauchy formula}
\begin{eqnarray}
\label{A1}
C_k\left({\sf B}\right) C_k\left({\sf D}\right) 
&=& C_k\left({\sf B D}\right)
\end{eqnarray}
from which immediately follows 
\begin{eqnarray}
\label{A1c}
C_k\left({\sf B}^{-1}\right) &=& C_k\left({\sf B}\right)^{-1}\ ,
\end{eqnarray}
the
{\it Laplace expansion}
\begin{eqnarray}
\label{A2}
C_k\left({\sf B}\right) C^{\; n-k}\left({\sf B}\right)\ =\
C^{\; n-k}\left({\sf B}\right) C_k\left({\sf B}\right)&=&\nonumber\\[0.3cm]
C_k\left({\sf B}\right) C_{n-k}\left({\sf B}\right)^\star\ =\
C_{n-k}\left({\sf B}\right)^\star C_k\left({\sf B}\right)
&=&\det {\sf B}\ {\bf 1}_{n\choose k}\ ,
\end{eqnarray}
{\it Jacobi's theorem} (a consequence of the eqs.\ (\ref{A2}) and
(\ref{A1c}))
\begin{eqnarray}
\label{A3}
C_k\left({\sf B}^{-1}\right) &=&
\frac{1}{\det {\sf B}}\ C^{\; n-k}\left({\sf B}\right)\ =\
\frac{1}{\det {\sf B}}\ C_{n-k}\left({\sf B}\right)^\star\ ,
\end{eqnarray}
and the
{\it Sylvester-Franke theorem}
\begin{eqnarray}
\label{A4}
\det C_k\left({\sf B}\right)&=&\left(\det {\sf B}\right)^{n-1\choose k-1}\ .
\end{eqnarray}

Compound matrices are treated in a number of references.
A comprehensive discussion of compound matrices can be found in 
\cite{wedd}, Chap.\ V, pp.\ 63-87, \cite{aitk}, Chap.\ V,
pp.\ 90-110, and, in a modern treatment, in \cite{fied}, Chap.\ 6,
pp.\ 142-155. More algebraically oriented modern treatments can
be found in \cite{marc1}, Part I, Chap.\ 2, Sect.\ 2.4, pp.\ 116-159,
Part II, Chap.\ 4, pp.\ 1-164 (very thorough),
\cite{jaco}, Chap.\ 7, Sect.\ 7.2, pp.\ 411-420, and
\cite{cohn}, Vol.\ 3, Chap.\ 2, Sect.\ 2.4, pp.\ 58-68. Concise
reviews of the properties of compound matrices are given
in \cite{bout,prel}. Also note \cite{vivi} and \cite{barn}.\\

\setcounter{section}{2}
\section*{Appendix B}
\setcounter{equation}{0}

Let ${\sf B}$ be a $3\times 3$ matrix. Then, the following identities apply. 
\begin{eqnarray}
\label{B1}
{\rm adj}\; {\sf B}&=&{\sf B}^2\ -\ 
{\sf B}\ {\rm tr}\; {\sf B}\ +\ \frac{1}{2}\ 
\left({\rm tr}\; {\sf B}\right)^2\ {\bf 1}_3\ -\ \frac{1}{2}\ 
{\rm tr}\left( {\sf B}^2\right)\ {\bf 1}_3\\[0.3cm]
\label{B2}
{\rm tr}\left( {\rm adj}\; {\sf B}\right)&=&
\frac{1}{2}\ \left({\rm tr}\; {\sf B}\right)^2\ -\ 
\frac{1}{2}\ {\rm tr}\left( {\sf B}^2\right)
\end{eqnarray}
Eq.\ (\ref{B1}) can be derived by means of the Cayley-Hamilton theorem 
(cf., e.g, \cite{spen1}, Subsect.\ 2.4, p.\ 264, eq.\ (2.4.7),
\cite{spen2}, Sect.\ 7, p.\ 154, eq.\ (29)).\\

\setcounter{section}{3}
\section*{Appendix C}
\setcounter{equation}{0}

In this Appendix we want to rederive eq.\ (\ref{QE8})
starting from a (Euclidean space-time) version
of the eqs.\ (\ref{M1a})-(\ref{M2}) on a finite lattice
with $k$ sites. The equations (\ref{M1c}), (\ref{M2})
then read
\begin{eqnarray}
\label{C1}
G[\phi] &=& W[J]
- \sum_{l=1}^k\ J_l \phi_l\ ,\\[0.3cm]
\label{C2}
J_l&=&-\ \frac{\partial G}{\partial\phi_l}\ .
\end{eqnarray}
In analogy to the eqs.\ (\ref{QD21}), (\ref{QD23}) 
we define (${\sf B}$ is a symmetric $k\times k$ matrix)
\begin{eqnarray}
\label{C3}
G_q&=&G_q[\phi]\ =\ 
-\frac{1}{2}\ \sum_{l,m=1}^k {\sf B}_{lm} \phi_l \phi_m
\ =\ -\frac{1}{2}\ \phi {\sf B} \phi\ ,\\[0.3cm]
\label{C4}
W_q&=&W_q[J]\ =\ 
\frac{1}{2}\ \sum_{l,m=1}^k 
\left({\sf B}^{-1}\right)_{lm} J_l J_m\ =\
\frac{1}{2}\ J {\sf B}^{-1} J\ .
\end{eqnarray}
Now we assume that $G$, $W$ depend on $\phi$, $J$ only as
functions of $G_q[\phi]$, $W_q[J]$, respectively,
and, in addition, that the identity $G = W$ holds.
Then, in analogy to the eqs.\ (\ref{QE3})-(\ref{QE5})
the eqs.\ (\ref{C1}), (\ref{C2}) read
\begin{eqnarray}
\label{C5}
G(G_q) &=& G(W_q)
- \sum_{l=1}^k\ J_l \phi_l\ ,\\[0.3cm]
\label{C6}
J_l &=&-\
\frac{\partial G(G_q)}{\partial\phi_l}\ =\ 
-\ G^{\;\prime}(G_q)\ \left(\phi {\sf B}\right)_l\ .
\end{eqnarray}
Here, again
\begin{eqnarray}
\label{C7}
G^{\;\prime}(G_q)&=&\frac{\partial G(G_q)}{\partial G_q}\ .
\end{eqnarray}
In view of eq.\ (\ref{C6}) it holds
\begin{eqnarray}
\label{C8}
W_q&=&- G_q\ \left[G^{\;\prime}(G_q)\right]^2 \ .
\end{eqnarray}
Taking into account the eqs.\ (\ref{C6}), (\ref{C8}),
eq.\ (\ref{C5}) can be written as
\begin{eqnarray}
\label{C9}
G(s)&=&G\left(-s \left[G^{\;\prime}(s)\right]^2\right)\ +\
2 s\ G^{\;\prime}(s)\ ,\ \ s\ =\ G_q\ ,
\end{eqnarray}
and this equation completely agrees with eq.\  (\ref{QE8}).\\

\end{document}